\documentclass[11pt,twoside]{article}
\usepackage{latexsym}
\usepackage{amssymb,amsbsy,amsmath,amsfonts,amssymb,amscd}
\usepackage{epsfig, graphicx}
\usepackage{hyperref}
\newcommand{\fer}[1]{(\ref{#1})}

\newcommand{\commentout}[1]{}
\newcommand{\R}{\mathbb{R}}

\newcommand {\lb} {\lambda}
\newcommand {\Chi} {{\bf \raise 2pt \hbox{$\chi$}} }
\newcommand {\f}   {\frac}
\newcommand {\p}   {\partial}
\newcommand{\dis}{\displaystyle}
\newcommand {\proof} {\noindent {\bf Proof}. }
\newcommand{\beq}{\begin{equation}}
\newcommand{\eeq}{\end{equation}}
\newcommand{\bea} {\begin{array}{rl}}
\newcommand{\eea} {\end{array}}
\newcommand{\bepa}{\left\{ \begin{array}{l}}
\newcommand{\eepa} {\end{array}\right.}
\newtheorem{theorem}{Theorem}[section]
\newtheorem{lemma}[theorem]{Lemma}
\newtheorem{definition}[theorem]{Definition}
\newtheorem{remark}[theorem]{Remark}
\newtheorem{proposition}[theorem]{Proposition}
\newtheorem{corollary}[theorem]{Corollary}

\numberwithin{equation}{section}

\newcommand{\qed}{{ \hfill
                       {\unskip\kern 6pt\penalty 500 \raise -2pt\hbox{\vrule\vbox to 6pt{\hrule width 6pt
                       \vfill\hrule}\vrule} \par}   }}

\title{\Large \bf Analysis of Nonlinear Noisy Integrate\&Fire Neuron \\ Models: blow-up and steady states}

\author{Mar\' {\i}a J. C\'aceres
\thanks{Departamento de Matem\'atica Aplicada,
Universidad de Granada, E-18071 Granada, Spain.
\texttt{caceresg@ugr.es}} \and Jos\'e A. Carrillo \thanks{
 ICREA and Departament de Matem\`atiques,
  Universitat Aut\`onoma de Barcelona
         E-08193 - Bellaterra, Spain.
\texttt{carrillo@mat.uab.cat}}
\and Beno\^ \i t Perthame\thanks{ 1-
Laboratoire Jacques-Louis Lions, UPMC, CNRS UMR 7598 and INRIA-Bang,
F-75005, Paris, France 2- Institut
Universitaire de France. \texttt{benoit.perthame@upmc.fr}}
}

\date{\today}

\begin{document}
\maketitle
\pagestyle{plain}
\pagenumbering{arabic}

\begin{abstract}
Nonlinear Noisy Leaky Integrate and Fire (NNLIF) models for
neurons networks can be written as Fokker-Planck-Kolmogorov
equations on the probability density of neurons, the main
parameters in the model being the connectivity of the network and
the noise. We analyse several aspects of the NNLIF model: the
number of steady states, a priori estimates, blow-up issues and
convergence toward equilibrium in the linear case. In particular,
for excitatory networks, blow-up always occurs for initial data
concentrated close to the firing potential. These results show how
critical is the balance between noise and excitatory/inhibitory
interactions to the connectivity parameter.
\end{abstract}

\\
\\
{\bf AMS Class. No}:  35K60, 82C31, 92B20

\section{Introduction}
\label{sec:intro}

The classical description of the dynamics of a large set of
neurons is based on deterministic/stochastic differential systems
for the excitatory-inhibitory neuron network \cite{lapicque,T}.
One of the most classical models is the so-called noisy leaky
integrate and fire (NLIF) model. Here, the dynamical behavior of
the ensemble of neurons is encoded in a stochastic differential
equation for the evolution in time of the averaged action
potential of the membrane $v(t)$ of a typical neuron
representative of the network. The neurons relax towards their
resting potential $V_L$ in the absence of any interaction. All the
interactions of the neuron with the network are modelled by an
incoming synaptic current $I(t)$. More precisely, the evolution of
the action potential follows, see \cite{BrHa,brunel,RBW,CBGW}
\begin{equation}\label{lif}
C_m \frac{dV}{dt} = -g_L (V-V_L) + I(t)
\end{equation}
where $C_m$ is the capacitance of the membrane and $g_L$ is the
leak conductance, normally taken to be constants with
$\tau_m=g_L/C_m\approx 2ms$ being the typical relaxation time of
the potential towards the leak reversal (resting) potential
$V_L\approx -70mV$. Here, the synaptic current takes the form of
a stochastic process given by:
\begin{equation}\label{synapcur}
I(t) = J_E \sum_{i=1}^{C_E} \sum_j \delta(t-t_{Ej}^i) - J_I
\sum_{i=1}^{C_I} \sum_j \delta(t-t_{Ij}^i)\,,
\end{equation}
where $\delta$ is the Dirac Delta at 0. Here, $J_E$ and $J_I$ are
the strength of the synapses, $C_E$ and $C_I$ are the total number
of presynaptic neurons and $t_{Ej}^i$ and $t_{Ij}^i$ are the the
times of the $j^{th}$-spike coming from the $i^{th}$-presynaptic
neuron for excitatory and inhibitory neurons respectively. The
stochastic character is embedded in the distribution of the spike
times of neurons. Actually, each neuron is assumed to spike
according to a stationary Poisson process with constant
probability of emitting a spike per unit time $\nu$. Moreover, all
these processes are assumed to be independent between neurons.
With these assumptions the average value of the current and its
variance are given by $\mu_C=b \nu$ with $b=C_E J_E - C_I J_I$ and
$\sigma_C^2=(C_E J_E^2+C_I J_I^2)\nu$. We will say that the
network is average-excitatory (average-inhibitory resp.) if $b>0$
($b<0$ resp.).

Being the discrete Poisson processes still very difficult to
analyze, many authors in the literature \cite{brunel,BrHa,RBW,mg}
have adopted the diffusion approximation where the synaptic
current is approximated by a continuous in time stochastic process
of Ornstein-Uhlenbeck type with the same mean and variance as the
Poissonian spike-train process. More precisely, we approximate
$I(t)$ in \eqref{synapcur} as
$$
I(t)\,dt\approx \mu_c \,dt + \sigma_C\, dB_t
$$
where $B_t$ is the standard Brownian motion. We refer to the work
\cite{RBW} for a nice review and discussion of the diffusion
approximation which becomes exact in the infinitely large network
limit, if the synaptic efficacies $J_E$ and $J_I$ are scaled
appropriately with the network sizes $C_E$ and $C_I$.

Finally, another important ingredient in the modelling comes from
the fact that neurons only fire when their voltage reaches certain
threshold value called the threshold or firing voltage $V_F\approx
-50mV$. Once this voltage is attained, they discharge themselves,
sending a spike signal over the network. We assume that they
instantaneously relax toward a reset value of the voltage
$V_R\approx -60mV$. This is fundamental for the interactions with
the network that may help increase their action potential up to
the maximum level (excitatory synapses), or decrease it for
inhibitory synapses. Choosing our voltage and time units in such a
way that $C_m=g_L=1$, we can summarize our approximation to the
stochastic differential equation model \eqref{lif} as the
evolution given by
\begin{equation}\label{lifdef}
dV = (-V +V_L +\mu_c) \,dt + \sigma_C\, dB_t
\end{equation}
for $V\leq V_F$ with the jump process: $V(t_o^+)=V_R$ whenever at
$t_0$ the voltage achieves the threshold value $V(t_o^-)=V_F$;
with $V_L<V_R<V_F$. Finally, we have to specify the probability of
firing per unit time of the Poissonian spike train $\nu$. This is
the so-called firing rate and it should be self-consistently
computed from a fully coupled network together with some external
stimuli. Therefore, the firing rate is computed as $\nu =
\nu_{ext} + N(t)$ where $N(t)$ is the mean firing rate of the
network. The value of $N(t)$ is then computed as the flux of
neurons across the threshold or firing voltage $V_F$. We finally
refer to \cite{G} for a nice brief introduction to this subject.

Coming back to the diffusion approximation in \eqref{lifdef}, we
can write a partial differential equation for the evolution of the
probability density $p(v,t)\geq 0$ of finding neurons at a voltage
$v\in (-\infty,V_F]$ at a time $t\geq 0$. Standard Ito's rule
gives the backward Kolmogorov or Fokker-Planck equation
\begin{equation}
\label{eq:nif1} \frac{\partial p}{\partial t} (v,t) +
\frac{\partial}{\partial v} \left[h\big(v, N(t)\big) p(v,t)\right]
- a\big(N(t)\big) \frac{\p^2 p}{\p  v^2} (v,t) =  \delta(v-V_R)
N(t), \qquad v\leq V_F \, ,
\end{equation}
with $h(v,N(t))=-v +V_L + \mu_c$ and $a(N)=\sigma^2_C/2$. We have
the presence of a source term in the right-hand side due to all
neurons that at time $t\geq 0$ fired, sent the signal on the
network and then, their voltage was immediately reset to voltage
$V_R$. Moreover, no neuron should have the firing voltage due to
the instantaneous discharge of the neurons to reset value $V_R$,
then we complement \eqref{eq:nif1} with Dirichlet and initial
boundary conditions
\begin{equation} \label{eq:nif2}
p(V_F,t)=0, \qquad p (-\infty,t)=0, \qquad p(v,0)=p^0(v)\, .
\end{equation}
Equation \eqref{eq:nif1} should be the evolution of a probability
density, therefore
$$
\int_{-\infty}^{V_F} p(v,t)\,dv = \int_{-\infty}^{V_F} p^0(v)\,dv
=1
$$
for all $t\geq 0$. Formally, this conservation should come from
integrating \eqref{eq:nif1} and using the boundary conditions
\eqref{eq:nif2}. It is straightforward to check that this
conservation for smooth solutions is equivalent to characterize
the mean firing rate for the network $N(t)$ as the flux of neurons
at the firing rate voltage. More precisely, the mean firing rate
$N(t)$ is implicitly given by
\begin{equation} \label{eq:nif3}
N(t):= -a\big(N(t)\big) \frac{\p p}{\p v} (V_F,t) \geq 0 \,.
\end{equation}
Here, the right-hand side is nonnegative since $p\geq 0$ over the
interval $[-\infty,V_F]$ and thus, $\frac{\p p}{\p v} (V_F,t) \leq
0$. In particular this imposes a limitation on the growth of the
function $N\mapsto a(N)$ such that \eqref{eq:nif3} has a unique
solution $N$.

The above Fokker-Planck equation has been widely used in
neurosciences. Often the authors prefer to write it in an
equivalent but less singular form.  To avoid the Dirac delta in
the right hand side, one can also set the same equation on
$(-\infty, V_R)\cup(V_R, V_F]$ and introduce the jump condition
$$
p(V_R^-,t)=p(V_R^+,t), \qquad \frac{\p}{\p v} p(V_R^-,t)-\frac{\p}{\p v} p(V_R^+,t)=N(t).
$$
This is completely transparent in our analysis which relates on a
weak form that applies to both settings.

Finally, let us choose a new voltage variable by translating it
with the factor $V_L+b \nu_{ext}$ while, for the
sake of clarity, keeping the notation for
the rest of values of the potentials involved $V_R<V_F$.
In these new variables, the drift and diffusion
coefficients are of the form
\begin{equation}
\label{as:nifbas}
h(v, N)= -v+bN, \qquad a(N)= a_0+a_1 N
\end{equation}
where $b>0$ for excitatory-average networks and $b<0$ for
inhibitory-average networks, $a_0>0$ and $a_1\geq 0$. Some results
in this work can be obtained for some more general drift and
diffusion coefficients. The precise assumptions will be specified
on each result. Periodic solutions have been numerically reported
and analysed in the case of the Fokker-Planck equation for
uncoupled neurons in \cite{NKKRC10,NKKZRC10}. Also, they study the
stationary solutions for fully coupled networks obtaining and
solving numerically the implicit relation that the firing rate $N$
has to satisfy, see Section 3 for more details.

There are several other routes towards modeling of spiking neurons
that are related to ours and that have been used in neurosciences,
see \cite{GK}. Among them are the deterministic I\&F models with
adaptation which are known for fitting well experimental data
\cite{BrGe}. In this case it is known that in the quadratic (or
merely superlinear) case, the model can blow-up
\cite{Touboul_AQIF}. One can also introduce gating variables in
neuron networks and this leads to a kinetic equation, see
\cite{CCTa} and the references therein. Another method consists in
coding the information in the distribution of time elapsed between
discharges \cite{PPCV, PPD}, this leads to nonlinear models that
exhibit naturally periodic activity, but blow-up has not been
reported.

In this work we will analyse certain properties of the solutions
to \eqref{eq:nif1}-\eqref{eq:nif2} with the nonlinear term due to
the coupling of the mean firing rate given by \eqref{eq:nif3}.
Next section is devoted to a finite time blow-up of weak solutions
for \eqref{eq:nif1}--\eqref{eq:nif3}. In short, we show that
whenever the value of $b>0$ is, we can find suitable initial data
concentrated enough at the firing rate such that the defined weak
solutions do not exist for all times. This implies that this model
encodes complicated dynamics. As long as the solution exists in
the sense specified in Section \ref{sec:estimates}, we can get
apriori estimates on the $L^1_{loc}$-norm of the firing rate.
Section \ref{sec:stst} deals with the stationary states of
\eqref{eq:nif1}--\eqref{eq:nif3}. We can show that there are
unique stationary states for $b\leq 0$ and $a$ constant but for
$b>0$ different cases may happen: one, two or no stationary states
depending on how large $b$ is. In Section \ref{sec:linear}, we
discuss the linear problem $b=0$ with $a$ constant for which the
general relative entropy principle applies implying the
exponential convergence towards equilibrium. Finally, Section
\ref{sec:num} is devoted to some numerical simulations of the
model showing some of the results here and getting some
conjectures about the nonlinear stability of the found stationary
states.


\section{Finite time blow-up and apriori estimates for weak solutions}
\label{sec:estimates}

Since we study a nonlinear version of the backward Kolmogorov or
Fokker-Planck equation \eqref{eq:nif1}, we start with the notion
of solution:

\begin{definition}\label{defweak}
We say that a pair of nonnegative functions $(p,N)$ with $p\in
L^\infty \big(\R^+; L^1_+(-\infty, V_F)\big)$, $N\in L^1_{\rm loc,
+}(\R^+)$ is a weak solution of \eqref{eq:nif1}--\eqref{as:nifbas}
if for any test function $\phi(v,t) \in C^\infty
((-\infty,V_F]\times[0,T])$ such that $\f{\p^2 \phi}{\p v^2} $, $v
\frac{\partial\phi}{\partial v} \in L^
\infty((-\infty,V_F)\times(0,T))$, we have
\begin{align}\label{eq:defweak}
\int_0^T \int_{-\infty}^{V_F} p(v,t)  \left[ -\f{\p \phi}{\p t}
-\f{\p \phi}{\p v} h(v,N)  -a \f{\p^2 \phi}{\p v^2}  \right] dv\,
dt = &\,\int_0^T N(t) [\phi(V_R,t)- \phi(V_F,t)] \, dt
\\
& + \int_{-\infty}^{V_F} p^0(v) \phi(0,v) \,dv-
\int_{-\infty}^{V_F} p(v,T) \phi(T,v) \,dv.\nonumber
\end{align}
\end{definition}

Let us remark that the growth condition on the test function
together with the assumption \eqref{as:nifbas} imply that the term
involving $h(v,N)$ makes sense. By choosing test functions of the
form $\psi(t)\phi(v)$, this formulation is equivalent to say that
for all $\phi(v) \in C^\infty ((-\infty,V_F])$ such that $v
\frac{\partial\phi}{\partial v} \in L^ \infty((-\infty,V_F))$, we
have that
\begin{align}\label{eq:defweakp}
\frac{d}{dt} \int_{-\infty}^{V_F}\phi(v) p(v,t) dv =
\int_{-\infty}^{V_F}   \left[ \f{\p \phi}{\p v} h(v,N) + a \f{\p^2
\phi}{\p v^2}  \right]\;  p(v,t) dv +N(t) [\phi(V_R,t)-
\phi(V_F,t)]
\end{align}
holds in the distributional sense. It is trivial to check that
weak solutions conserve the mass of the initial data by choosing
$\phi=1$ in \eqref{eq:defweakp}, and thus,
\begin{equation}\label{massconserv}
\int_{-\infty}^{V_F} p(v,t)\,dv = \int_{-\infty}^{V_F} p^0(v)\,dv
=1 \,.
\end{equation}
The first result we show is that global-in-time weak solutions of
\eqref{eq:nif1}--\eqref{eq:nif3} do not exist for all initial data
in the case of an average-excitatory network. This result holds
with less stringent hypotheses on the coefficients than in
\eqref{as:nifbas} with an analogous notion of weak solution as in
Definition \ref{eq:defweak}.

\begin{theorem} [Blow-up]
Assume that the drift and diffusion coefficients satisfy
\begin{equation}\label{as:nif1}
h(v,N)+v \geq b N \qquad \mbox{and} \qquad a(N) \geq a_m > 0,
\end{equation}
for all $-\infty < v \leq V_F$ and all $N\geq 0$, and let us
consider the average-excitatory network where $b>0$. If the
initial data is concentrated enough around $v=V_F$, in the sense
that
$$
\int_{-\infty}^{V_F} e^{\mu v} p^0(v)\, dv
$$
is large enough with $\mu>\max(\frac{V_F}{a_m},\frac1{b})$, then
there are no global-in-time weak solutions to
\eqref{eq:nif1}--\eqref{eq:nif3}. \label{th:bu}\end{theorem}

\proof We choose a multiplier $\phi(v)=e^{\mu v}$ with $\mu >0$
and define the number
$$
\lb = \f{ \phi(V_F) - \phi(V_R) }{b\mu} >0
$$
by hypotheses. For a weak solution according to \fer{eq:defweak},
we find from \eqref{eq:defweakp} that
\begin{align}\label{boundbelow}
\f{d}{dt} \int_{-\infty}^{V_F} \phi(v) p(v,t) \,dv & \geq \mu
\int_{-\infty}^{V_F} (b N(t)-v)\phi(v) p(v,t) \,dv + \mu^2
a_m\int_{-\infty}^{V_F} \phi(v) p(v,t) \,dv - \lb b \mu N(t)
\nonumber\\
& \geq   \mu \int_{-\infty}^{V_F} \phi(v) p(v,t) \,dv \; [ b N(t)+
\mu a_m -V_F  ] - \lb \mu b N(t)
\end{align}
where \eqref{as:nif1} and the fact that $v\in (-\infty,V_F)$ was
used. Let us now choose $\mu$ large enough such that $\mu a_m -V_F
>0$ according to our hypotheses and denote
$$
M_\mu (t) =\int_{-\infty}^{V_F} \phi(v) p(v,t)\,dv\, ,
$$
which satisfies
$$
\frac{d}{dt} M_\mu (t)  \geq b\mu N(t) [M_\mu (t) -\lb].
$$
If initially $M_\mu(0) \geq \lb$ and using Gronwall's Lemma since
$N(t)\geq 0$, we have that $M_\mu(t)\geq \lb$, for all $t\geq 0$,
and back to \eqref{boundbelow} we find
$$
\f{d}{dt} \int_{-\infty}^{V_F} \phi(v) p(v,t)\,dv \geq \mu (\mu
a_m -V_F) \int_{-\infty}^{V_F} \phi(v) p(v,t) \,dv
$$
which in turn implies,
$$
\int_{-\infty}^{V_F} \phi(v) p(v,t) \,dv \geq e^{\mu (\mu a_m
-V_F) t} \int_{-\infty}^{V_F} \phi(v) p^0(v) \,dv.
$$
On the other hand, since $p(v,t)$ preserves the mass, see
\eqref{massconserv}, and $\mu>0$ then
$$
\int_{-\infty}^{V_F} \phi(v) p(v,t) \,dv \leq e^{\mu V_F},
$$
leading to a contradiction.

It remains to show that the set of initial data satisfying the
size condition in the statement is not empty. To verify this, we
can approximate as much as we want by smooth initial probability
densities an initial Dirac mass at $V_F$ which gives the condition
$$
e^{\mu V_F}  \geq \lb = \f{ e^{\mu V_F}   - e^{\mu V_R}   }{b\mu}
\quad \text{together with} \;  \mu a_m > V_F.
$$
This can be equivalently written as
$$
\mu \geq \f{ 1 - e^{-\mu (V_F-V_R)}  }{b} \quad \text{and }\;
\mu
> \f{V_F}{a_m}.
$$
Choosing $\mu$ large enough, these
conditions are obviously fulfilled. \qed

\bigskip

As usual for this type of blow-up result similar in spirit to the
classical Keller-Segel model for chemotaxis \cite{BDP,CPZ}, the
proof only ensures that solutions for those initial data do not
exist beyond a finite maximal time of existence. It does not
characterize the nature of the first singularity which occurs. It
implies that either the decay at infinity is false, although not
probable, implying that the time evolution of probability
densities ceases to be tight, or the function $N(t)$ may become a
singular measure in finite time instead of being an
$L^1_{loc}(\R^+)$ function. Actually, in the numerical
computations shown in Section \ref{sec:linear}, we observe a blow-up in the value
of the mean firing rate in finite time. This will need a
modification of the notion of solution introduced in Definition
\ref{defweak}.

Nevertheless, it is possible to obtain some a priori bounds with
the help of appropriate choices of the test function $\phi$ in
\eqref{eq:defweak}. Some of these choices are not allowed due to
the growth at $-\infty$ of the test functions. We will say that a
weak solution is fast-decaying at $-\infty$ if they are weak
solutions in the sense of Definition \ref{defweak} and the weak
formulation in \eqref{eq:defweakp} holds for all test functions
growing algebraically in $v$.

\begin{lemma} [A priori estimates]
Assume \eqref{as:nifbas} on the drift and diffusion coefficients and that $(p,N)$
is a global-in-time solution of \eqref{eq:nif1}--\eqref{eq:nif3}
in the sense of Definition {\rm \ref{defweak}} fast decaying at $-\infty$, then the following
apriori estimates hold:
\begin{itemize}
\item[(i)] If $b \geq  V_F - V_R$, then
$$
\int_{-\infty}^{V_F} (V_F-v) p(v,t) dv \leq \max \left(V_F,\int_{-\infty}^{V_F} (V_F-v) p^0(v) dv)\right),
$$
$$
(b-V_F+V_R) \int_0^T N(t) dt \leq V_F T+\int_{-\infty}^{V_F} (V_F-v) p^0(v) dv,
$$
\item[(ii)] If $b< V_F - V_R$ then
$$
\int_{-\infty}^{V_F} (V_F-v) p(v,t) dv \geq \min
\left(V_F,\int_{-\infty}^{V_F} (V_F-v) p^0(v) dv)\right) .
$$
Moreover, if in addition $a$ is constant then
$$
\int_0^T N(t)dt \leq (1+T) C(b, V_F-V_R,a).
$$
\end{itemize}
\label{lm:apriori}\end{lemma}

\proof  With our decay assumption at $-\infty$, we may use the
test function $\phi(v) =V_F- v \geq 0$. Then \eqref{eq:defweakp}
gives
$$
\frac{d}{dt} \int_{-\infty}^{V_F} \phi(v) p(v,t) dv = \int_{-\infty}^{V_F} [v- bN(t)] p(v,t) dv + N(t) (V_F-V_R).
$$
This is also written as
\begin{equation} \label{weakv}
\frac{d}{dt} \int_{-\infty}^{V_F} \phi(v) p(v,t) dv+
\int_{-\infty}^{V_F} \phi(v) p(v,t) dv =V_F- N(t) \; [b -
(V_F-V_R)].
\end{equation}

To prove (i), with our condition on $b$ the term in $N(t)$ is nonpositive and both results follow after  integration in time.
\\

To prove (ii), we first  use again \eqref{weakv} and, because the term in $N(t)$ is nonnegative, we find the first result.  Then, we use a truncation function $\phi(v)\in C^2$ such that
$$
\phi(V_F)=1, \qquad \phi(v)=0 \text{ for } v \leq V_R, \qquad \phi'(v) \geq 0.
$$
Equation \eqref{eq:defweakp} gives
$$
 \f{d}{dt} \int_{V_R}^{V_F} \phi(v) p(v,t)\, dv +N(t)= \int_{V_R}^{V_F} \phi'(v) \big(-v+b N(t)\big) p(v,t)\,dv + a \int_{V_R}^{V_F} \phi''(v) p(v,t)\,dv\,.
$$
Except regularity at $v=V_R$, we can have in mind $\phi'(v)
=1/(V_F-V_R)$ for $v>V_R$, and we can be as close as we want of
this choice by paying a large second derivative of $\phi$.  So
with the parameter
$$
\delta =\frac{1}{2} \left[ 1 - \frac{b}{V_F-V_R}\right]
$$
we may achieve with $C(\delta ) $ large
$$
 \f{d}{dt} \int_{V_R}^{V_F} \phi(v) p(v,t) \,dv + \delta N(t) \leq a\int_{V_R}^{V_F} \phi''(v) p(v,t)\,dv\leq C(\delta).
$$
And this leads directly to the result after integration in time.
\qed

\begin{corollary} Under the assumptions of Lemma
{\rm\ref{lm:apriori}} and assuming $v^2 p^0(v)\in
L^1(-\infty,V_F)$ and $0<b < V_F - V_R$, then the following
apriori estimates hold:
\begin{itemize}
\item[(i)] If additionally $a$ is constant, for all $t\geq 0$ we
have
$$
\int_{-\infty}^{V_F} v^2 p(v,t)\, dv \leq C(1+t)
$$

\item[(ii)] If additionally $-b\min \left(V_F,\int_{-\infty}^{V_F}
(V_F-v) p^0(v) dv)\right)  + a_1+bV_F
 + \frac{V_R^2-V_F^2}{2}\ \leq 0$, then
$$
\int_{-\infty}^{V_F} v^2 p(v,t)\, dv \leq
\max\left(a_0,\int_{-\infty}^{V_F} v^2 p^0(v,t)\, dv\right).
$$
\end{itemize}
\end{corollary}

\proof We use $\phi(v)=v^2/2$ as test function to get
\begin{align*}
 \f{d}{dt} \int_{-\infty}^{V_F} \frac{v^2}2 p(v,t)\, dv +& \int_{-\infty}^{V_F} v^2 p(v,t)\,dv = bN(t)\int_{-\infty}^{V_F} v p(v,t)\,dv + a(N(t))
 + N(t) \frac{V_R^2-V_F^2}{2} \\
&= bN(t) \int_{-\infty}^{V_F} (v-V_F)  p(v,t)\,dv+ a(N(t))
 + N(t)\left[bV_F+  \frac{V_R^2-V_F^2}{2}\right] \\
 &\leq  a_0 + N(t)\left[-b\min \left(V_F,\int_{-\infty}^{V_F} (V_F-v) p^0(v) dv)\right)  + a_1+bV_F
 + \frac{V_R^2-V_F^2}{2}\right]
\end{align*}
thanks to  the first statement of Lemma \ref{lm:apriori} (ii).
\\

To prove (i), we just use the second statement of Lemma
\ref{lm:apriori} (ii) valid for $a$ constant which tells us that
the time integration of the right-hand side grows at most linearly
in time and so does $\int_{-\infty}^{V_F} v^2 p(v,t)\,dv$.
\\

To prove (ii), we just use that the bracket is nonpositive and the
results follows. \qed


\section{Steady states}
\label{sec:stst}


\subsection{Generalities}
This section is devoted to find all smooth stationary solutions of
the problem \eqref{eq:nif1}-\eqref{eq:nif3} in the particular
relevant case of a drift of the form $h(v)=V_0(N)-v $. Let us
search for continuous stationary solutions $p$ of \eqref{eq:nif1}
such that $p$ is $C^1$ regular except possibly at $V=V_R$ where it
is $Lipschitz$. Using the definition in \eqref{eq:defweakp}, we
are then allowed by a direct integration by parts in the second
derivative term of $p$ to deduce that $p$ satisfies
\begin{equation}
\frac{\partial}{\partial v}\left[
(v-V_0(N))p+a(N)\frac{\partial}{\partial v}p(v) + N H(v-V_R)
\right]=0 \label{stationary0}
\end{equation}
in the sense of distributions, with $H$ being the Heaviside
function, i.e., $H(u)=1$ for $u\geq 0$ and $H(u)=0$ for $u<0$.
Therefore, we conclude that
$$
(v-V_0(N))p +a(N) \f{\p p}{\p v} + N H(v-V_R)=C \, .
$$
The definition of $N$ in \eqref{eq:nif3} and the Dirichlet
boundary condition \eqref{eq:nif2} imply $C=0$ by evaluating this
expression at $v=V_F$. Using again the boundary condition
\eqref{eq:nif2}, $p(V_F)=0$, we may finally integrate again and
find that
$$
p(v)= \frac{N}{a(N)} e^{-\f{(v-V_0(N))^2}{2a}}  \int_v^{V_F}
e^{\f{(w-V_0(N))^2}{2a}} H[w-V_R] dw
$$
which can be rewritten, using the expression of the Heaviside
function, as
\begin{equation}
p(v)= \f{N}{a(N)} e^{-\f{(v-V_0(N))^2}{2a}}  \int_{\max(v,
V_R)}^{V_F} e^{\f{(w-V_0(N))^2}{2a}} \,dw.
\label{stationaryp}
\end{equation}
Moreover, the firing rate in the stationary state $N$ is
determined by the normalization condition \eqref{massconserv}, or
equivalently,
\begin{equation}
\f{a(N)}{N}= \int_{-\infty}^{V_F} \left[ e^{-\f{(v-V_0(N))^2}{2a}}
\; \int_{\max(v, V_R)}^{V_F} e^{\f{(w-V_0(N))^2}{2a}} dw \right]
dv\,.
\label{stationarymass}
\end{equation}
Summarizing all solutions with the above referred regularity of
the stationary problem \eqref{stationary0} are of the form in
\eqref{stationaryp} with $N$ being any positive solution to
\eqref{stationarymass}.

Let us first comment that in the linear case $V_0(N)=0$ and
$a(N)=a>0$, we then get a unique stationary state $p_\infty$ given
by the expression
\begin{equation}
p_\infty(v)= \f{N_\infty}{a} e^{-\f{v^2}{2a}} \int_{\max(v,
V_R)}^{V_F} e^{\f{w^2}{2a}} \,dw. \label{stationaryplinear}
\end{equation}
with $N_\infty$ the normalizing constant to unit mass over the
interval $(-\infty,V_F]$, as obtained in \cite{BrHa}.

The rest of this section is devoted to find conditions on the
parameters of the model clarifying the number of solutions to
\eqref{stationarymass}. With this aim, it is convenient to perform
a change of variables, and use new notations
\begin{equation}
\label{change-var} z=\f{v- V_0}{\sqrt{a}}, \quad
u=\f{w-V_0}{\sqrt{a}}, \qquad w_F=\frac{V_F-V_0}{\sqrt{a}}, \quad
w_R=\frac{V_R-V_0}{\sqrt{a}} ,
\end{equation}
where the $N$ dependency has been avoided to simplify notation.
Then, we can rewrite the previous integral (and thus the condition
for a steady state) as
\begin{equation} \label{condst}
\left\{
\begin{array}{l} \dis \f 1 N=I(N) ,
\\ [3mm]
I(N):= \dis \int_{-\infty}^{w_F} \left[  e^{-\f{z^2}{2}} \;
\int_{\max(z,w_R)}^{w_F} e^{\f{u^2}{2}} du  \right]  dz.
\end{array}\right.
\end{equation}
Another alternative form  of $I(N)$ follows from the change of
variables $s=(z-u)/2$ and $\tilde s=(z+u)/2$ to get
$$
I(N) =  2 \int_{-\infty}^0\int_{w_R+s}^{w_F+s} e^{-2s \tilde s} \
d\tilde s \ ds = -\int_{-\infty}^0 \frac{e^{-2 s^2}}{s} \left(
e^{-2\, s\, w_F}-e^{-2\, s\, w_R}\right) ds \, ,
$$
and consequently,
\begin{equation} \label{eq:alteri}
I(N) = \int_0^{\infty} \frac{e^{- s^2/2}}{s}\left( e^{ s\,
w_F}-e^{s\, w_R}\right) ds.
\end{equation}

\subsection{Case of $a(N)=a_0$.}

We are now ready to state our main result on steady states.
\begin{theorem} Assume $h(v,N)=b\,N-v$, $a(N)=a_0$ is constant and
$V_0=b\,N$.
\begin{itemize}
\item[i)] For $b<0$ and $b>0$ small enough there is a unique
steady state to \eqref{eq:nif1}-\eqref{eq:nif3}.

\item[ii)] Under either the condition
\begin{equation}
\label{as:existst1} 0< b < V_F-V_R\,,
\end{equation}
or the condition
\begin{equation} \label{as:existst2}
0< 2 a_0 b < (V_F-V_R)^2 V_R\,,
\end{equation}
then there exists at least one steady state solution to
\eqref{eq:nif1}-\eqref{eq:nif3}.

\item[iii)] If both \eqref{as:existst2} and $b > V_F-V_R$ hold,
then there are at least two steady states to
\eqref{eq:nif1}-\eqref{eq:nif3}.

\item[iv)] There is no steady state to
\eqref{eq:nif1}-\eqref{eq:nif3} under the high connectivity
condition
\begin{equation} \label{as:existst3}
b> \max(2(V_F-V_R),2V_F\,I(0)).
\end{equation}
\end{itemize}
\label{th:stationary}
\end{theorem}

\begin{remark}
It is natural to relate the absence of steady state for $b$ large
with blow-up of solutions. However, Theorem {\rm\ref{th:bu}} in
Section {\rm\ref{sec:estimates}} shows this is not the only
possible cause since the blow-up can happen for initial data
concentrated enough around $V_F$ independently of the value of
$b>0$. See also Section {\rm 5} for related numerical results.
\end{remark}

\begin{figure}
\epsfxsize=3.in \hbox to \hsize{\epsfbox{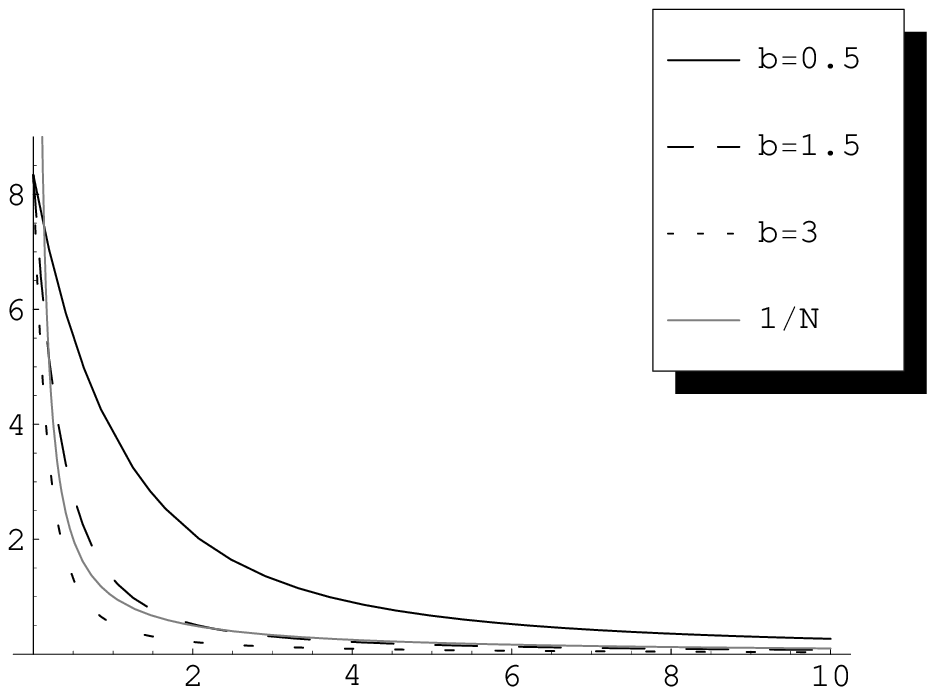}
\epsfxsize=3.in \hfil \epsfbox{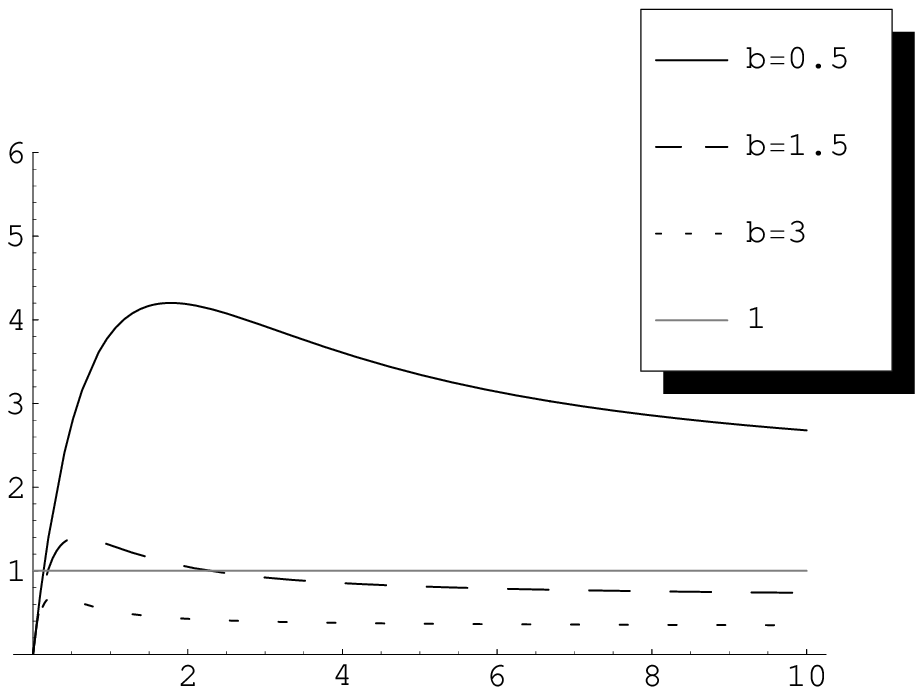}}

\caption{For several values of $b$, the function $I(N)$ in
\fer{condst} is plotted against the function $1/N$ (Left Figure)
and the function $NI(N)$ against the constant function $1$ (Right
Figure). Here $a\equiv 1$, $V_R = 1$, $V_F = 2$.} \label{fig:stst}
\end{figure}

\noindent {\bf Proof.} Let us first study properties of the
function $I(N)$. We first rewrite \eqref{eq:alteri} as
\begin{equation*}
I(N)=\int_0^{\infty} e^{- s^2/2} e^{-\frac{sbN}{\sqrt{a_0}}}
\frac{e^{\frac{s\, V_F}{\sqrt{a_0}}}-
 e^{\frac{s\, V_R}{\sqrt{a_0}}}}{s} \ ds.
\end{equation*}
A direct Taylor expansion implies that
\begin{equation}\label{Taylor}
\left|\frac{e^{\frac{s\, V_F}{\sqrt{a_0}}}-
 e^{\frac{s\, V_R}{\sqrt{a_0}}}}
{s} - \frac{V_F-V_R}{\sqrt{a_0}}\right| \leq C_0\, s\,
e^{\frac{s\,V_F}{\sqrt{a_0}}}
\end{equation}
for all $s\geq 0$. Then, a direct application of the dominated
convergence theorem and continuity theorems of integrals with
respect to parameters show that the function $I(N)$ is continuous
on $N$ on $[0,\infty)$. Moreover, the function $I(N)$ is
$C^\infty$ on $N$ since all their derivatives can be computed by
differentiating under the integral sign by direct application of
dominated convergence theorems and differentiation theorems of
integrals with respect to parameters. In particular,
\begin{equation*}
I'(N) =- \f{b}{\sqrt{a_0}}\; \int_0^{\infty} e^{- s^2/2} \left( e^{
s\, w_F}-e^{s\, w_R}\right) ds\, ,
\end{equation*}
and for all integers $k\geq 1$,
$$
I^{(k)}(N) =(-1)^k \left(\f{b}{\sqrt{a_0}}\right)^k\;
\int_0^{\infty} e^{- s^2/2} s^{k-1} \left( e^{ s\, w_F}-e^{s\,
w_R}\right) ds\, .
$$
As a consequence, we deduce:
\begin{itemize}
\item[1.] Case $b<0$: $I(N)$ is an increasing strictly convex function and thus
$$
\lim_{N\to\infty} I(N) = \infty \, .
$$

\item[2.] Case $b>0$: $I(N)$ is a decreasing convex function.
Also, it is obvious from the previous expansion \eqref{Taylor} and
dominated convergence theorem that
$$
\lim_{N\to\infty} I(N) = 0 \, .
$$
\end{itemize}
It is also useful to keep in mind that, thanks to the form of
$I(N)$ in \fer{condst},
\begin{equation}
\label{I0} I(0) \leq \sqrt{2\pi} [w_F(0)-w_R(0)]
e^{\max(w_R^2(0),w_F^2(0))/2} = \sqrt{2\pi} \frac{(V_F-V_R)}{a_0}
\exp\left\{\frac{\max(V_R^2,V_F^2)}{2a_0}\right\}<\infty \; .
\end{equation}
Now, let us show that for $b>0$, we have
\begin{equation} \label{eq:limnin}
\lim_{N\to \infty}N \; I(N)=\frac{V_F-V_R}{b}.
\end{equation}
Using \eqref{Taylor}, we deduce
$$
\left|NI(N)-N\frac{V_F-V_R}{\sqrt{a_0}}\int_0^{\infty} e^{- s^2/2}
e^{-\frac{sbN}{\sqrt{a_0}}}\ ds\right| \leq C_0 N \int_0^{\infty} s
\,e^{- s^2/2} e^{-\frac{sbN}{\sqrt{a_0}}} e^{\frac{s\,V_F}{\sqrt{a_0}}}\ ds.
$$
A direct application of dominated convergence theorem shows that
the right hand side converges to 0 as $N\to \infty$ since $sN
\exp(-\frac{sbN}{\sqrt{a_0}})$ is a bounded function uniform in $N$
and $s$. Thus, the computation of the limit is reduced to show
\begin{equation} \label{limit}
\lim_{N\to \infty} \, N\, \int_0^{\infty} e^{-
s^2/2-\frac{sbN}{\sqrt{a_0}}}\ ds=\frac{\sqrt{a_0}}{b}.
\end{equation}
With this aim, we rewrite the integral in terms of the
complementary error function defined as
$$
erfc(x):=\frac{2}{\sqrt{\pi}} \int_x^\infty e^{-t^2}\ dt,
$$
and then
$$
\int_0^\infty
e^{- s^2/2-\frac{sbN}{\sqrt{a_0}}}\ ds=
e^{\frac{b^2\, N^2}{2\, a_0}}
\int_0^\infty
e^{-(\frac{s}{\sqrt{2}}+\frac{bN}{\sqrt{2 a_0}})^2} \ ds=
\frac{\sqrt{\pi}}{\sqrt{2}}\,
e^{\frac{b^2\, N^2}{2\, a_0}}\,  erfc\left(\frac{bN}{\sqrt{2a_0}}\right).
$$
Finally, we can obtain the limit (\ref{limit}) using L'H\^opital's
rule
$$
\lim_{N\to \infty} \, N\int_0^{\infty} e^{-
s^2/2-\frac{sbN}{\sqrt{a_0}}}\ ds = \frac{\sqrt{\pi}}{\sqrt{2}}\,
\lim_{N\to \infty} \,
\frac{erfc(\frac{bN}{\sqrt{2a_0}})}{\frac{e^{-\frac{b^2\, N^2}{2\,
a_0}}}{N}} = \sqrt{2} \lim_{N\to \infty}
 \frac{-\frac{b}{\sqrt{2 a_0}}e^{-\frac{b^2N^2}{2a_0}}}
{ -\frac{b^2}{a_0}e^{-\frac{b^2N^2}{2a_0}}- \frac1{N^2}
e^{-\frac{b^2N^2}{2a_0}}}=\frac{\sqrt{a_0}}{b}.
$$
With this analysis of the function $I(N)$ we can now proof each of
the statements of Theorem \ref{th:stationary}:
\\
\\
\noindent {\bf Proof of i).} Let us start with the case $b<0$.
Here, the function $I(N)$ is increasing, starting at
$I(0)<\infty$ due to \eqref{I0} and such that
$$
\lim_{N\to\infty} I(N) = \infty \, .
$$
Therefore, it crosses to the function $1/N$ at a single point.

Now, for the case $b>0$ small, we first remark that similar
dominated convergence arguments as above show that both $I(N)$ and
$I'(N)$ are smooth functions of $b$. Moreover, it is simple to
realize that $I(N)$ is a decreasing function of the parameter $b$.
Now, choosing $0<b\leq b_*<(V_F-V_R)/2$, then $I(N)\geq I_*(N)$
for all $N\geq 0$ where $I_*(N)$ denotes the function associated
to the parameter $b_*$. Using the limit \eqref{eq:limnin}, we can
now infer the existence of $N_*>0$ depending only on $b_*$ such
that
$$
NI(N)\geq NI_*(N)> \frac{V_F-V_R}{2b_*} >1
$$
for all $N\geq N_*$. Therefore, by continuity of $NI(N)$ there are
solutions to $NI(N)=1$ and all possible crossings of $I(N)$ and
$1/N$ are on the interval $[0,N_*]$. We observe that both $I(N)$
and $I'(N)$ converge towards the constant function $I(0)>0$ and to
0 respectively, uniformly in the interval $[0,N_*]$ as $b\to 0$.
Therefore, for $b$ small $N\; I(N)$ is strictly increasing on the
interval $[0, N_*]$ and there is a unique solution to $N\;
I(N)=1$.
\\
\\
\noindent {\bf Proof of ii). Case of \eqref{as:existst1}.} The
claim that there are solutions to $NI(N)=1$ for $0<b<V_F-V_R$ is a
direct consequence of the continuity of $I(N)$, \eqref{I0} and
\eqref{eq:limnin}.
\\
\noindent {\bf Case of \eqref{as:existst2}.} We are going to prove
that $I(N) \geq 1/N$ for $\frac{2\, a_0}{(V_R-V_F)^2} < N <
\frac{V_R}{b}$, which concludes the existence of a steady state
since $I(0)<\infty$ due to \eqref{I0} implies that $I(N)<1/N$ for
small $N$. Condition \fer{as:existst2} only asserts that this
interval for $N$ is not empty. To do so, we show  that
\begin{equation*}
I(N)\geq
 \f{(V_R-V_F)^2}{2a} \quad \quad \mbox{for} \  N\in
\left[0, \frac{V_R}{b}\right]
\end{equation*}
which obviously concludes the desired inequality $I(N) \geq 1/N$
for the interval of $N$ under consideration.

The condition $\frac{V_R}{b}>N$ is equivalent to $w_R>0$,
therefore, using \eqref{change-var} and the expression for $I(N)$
in \eqref{condst}, we deduce
$$
I(N)\geq \int_{w_R}^{w_F} \left[
 e^{-\f{z^2}{2}} \;  \int_{\max(z,w_R)}^{w_F}
 e^{\f{u^2}{2}} du
\right] dz \ge \int_{w_R}^{w_F} \left[
 e^{-\f{z^2}{2}} \;  \int_z^{w_F}
 e^{\f{u^2}{2}} du
\right] dz\, .
$$
Since $z>0$ and $e^{\f{u^2}{2}}$ is an increasing function for
$u>0$, then $e^{\f{u^2}{2}}\geq e^{\f{z^2}{2}}$ on $[z,w_F]$, and
we conclude
$$
I(N)\ge \int_{w_R}^{w_F} \int_{z}^{w_F} \,du\, dz =
\f{(V_R-V_F)^2}{2a_0}.
$$
\\
\\
\noindent {\bf Proof of iii)}. Under the condition
\eqref{as:existst2}, we have shown in the previous point the
existence of an interval where $I(N)>I/N$. On one hand,
$I(0)<\infty$ in \eqref{I0} implies that $I(N)<I/N$ for $N$ small
and the condition $b>V_F-V_R$ implies that $I(N)<I/N$ for $N$
large enough due to the limit \eqref{eq:limnin}, thus there are at
least two crossings between $I(N)$ and $1/N$.
\\
\\
\noindent {\bf Proof of iv)}. Under assumption (\ref{as:existst3})
for $b$, it is easy to check that the following inequalities hold
\begin{equation}
\label{as:existst3-1} I(0)< 1/N \quad \mbox{for } N\le 2V_F/b
\end{equation} and
\begin{equation}\label{as:existst3-2}
\frac{V_F-V_R}{bN-V_F}<\frac{1}{N} \quad \mbox{for } N>2V_F/b.
\end{equation}
We consider $N$ such that $N>V_F/b$, this means that $w_F<0$. We
use the formula \fer{eq:alteri} for $I(N)$ and write the
inequalities
\begin{align*}
I(N) &< (w_F-w_R)  \int_0^\infty e^{-s^2/2} e^{sw_F} = (w_F-w_R)
e^{w_F^2/2}  \int_0^\infty e^{-(s-w_F)^2/2}
\\
&= (w_F-w_R) e^{w_F^2/2}  \int_{-w_F}^\infty e^{-s^2/2} \leq
(w_F-w_R) e^{w_F^2/2}  \int_{-w_F}^\infty \f{s}{|w_F|}e^{-s^2/2} =
\f{V_F-V_R}{\sqrt a \, |w_F|}
\end{align*}
where the mean-value theorem and $w_F <0$ were used. Then, we
conclude that
$$
I(N) < \f{V_F-V_R}{\sqrt a |w_F|}=\f{V_F-V_R}{bN-V_F} , \qquad \text{ for }ÊN> V_F/b.
$$
Therefore, using  Inequality (\ref{as:existst3-2}):
$$
I(N)  < \f 1 N  , \qquad \text{ for }N> 2 V_F/b
$$
and due to the fact that $I$ is decreasing and Inequality
(\ref{as:existst3-1}), we have $I(N) < I(0) < 1/N$, for $N \leq 2
V_F/b$. In this way, we have shown that for all $N$, $I(N)<1/N$
and consequently there is no steady state. \qed

\

\begin{remark}
The functions $I(N)$ and $1/N$ are depicted in
Figure {\rm\ref{fig:stst}} for the case $V_0(N)=bN$ and $a(N)=a_0$
illustrating the main result: steady states exist for small $b$
and do not exist for large $b$ while there is an intermediate
range of existence of two stationary states. The numerical plots
of the function $NI(N)$ might indicate that there are only three
possibilities: one stationary state, two stationary states and no
stationary state. However, we are not able to prove or disprove
the uniqueness of a maximum for the function $NI(N)$ eventually
giving this sharp result.
\end{remark}

\begin{remark}
The condition \eqref{as:existst2} can be improved by using one more term in the
series expansion of the exponentials inside the integral of the
expression of $I(N)$ in \eqref{eq:alteri}. More precisely,
if $w_F>w_R>0$,
 we use
$$
e^{sw_F}- e^{sw_R} = \sum_{n=0}^\infty \frac{s^n}{n!}
(w_F^n-w_R^n) \geq \sum_{n=0}^2 \frac{s^n}{n!} (w_F^n-w_R^n).
$$
In this way,  we get
$$
I(N) \ge \int_0^\infty e^{-s^2/2}\left(\frac{V_F-V_R}{\sqrt{a}} +
\frac{1}{2}\left( w_F^2 -w_R^2 \right)\, s \right) \ ds \ge
\frac{(V_F-V_R) \left( \sqrt{2\, \pi \, a} + (V_F-V_R)
\right)}{2\, a},
$$
since
$$
\int_0^\infty  e^{-s^2/2} \ ds = \sqrt{\f{\pi}{2}}, \quad
\int_0^\infty  e^{-s^2/2} \, s \ ds =1 \quad \mbox{and} \quad V_0<V_R.
$$
Then, condition \eqref{as:existst2} can be improved to
\begin{equation*} 2\, a \, b < V_R\,(V_F-V_R) \left(
\sqrt{2\, \pi \, a} + (V_F-V_R) \right). \label{as:existst2b}
\end{equation*}
\end{remark}

\begin{figure}
\epsfxsize=3.in \hbox to \hsize{\epsfbox{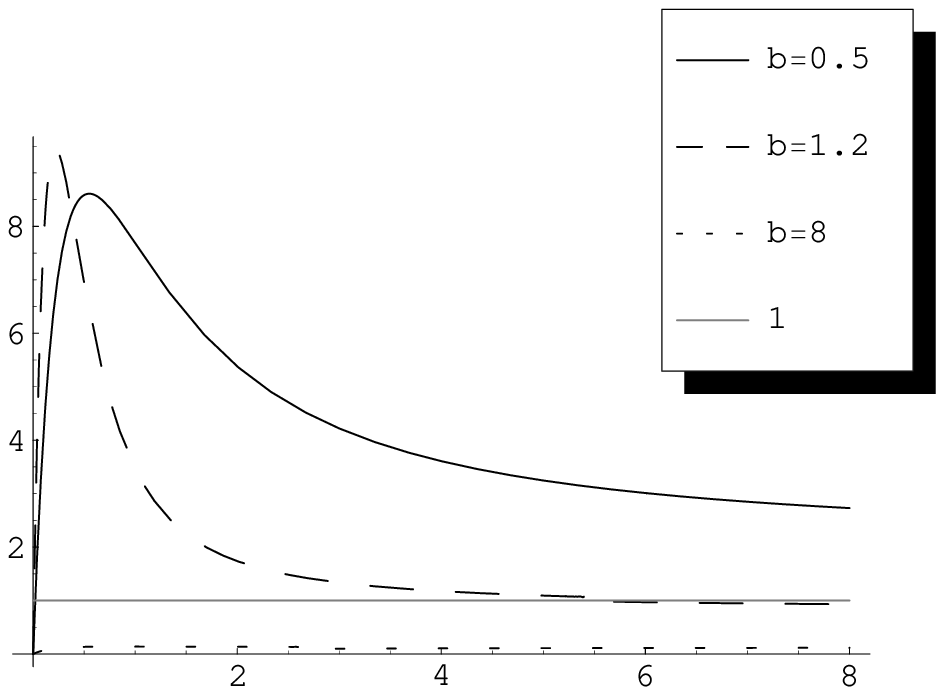}
\epsfxsize=3.in \hfil \epsfbox{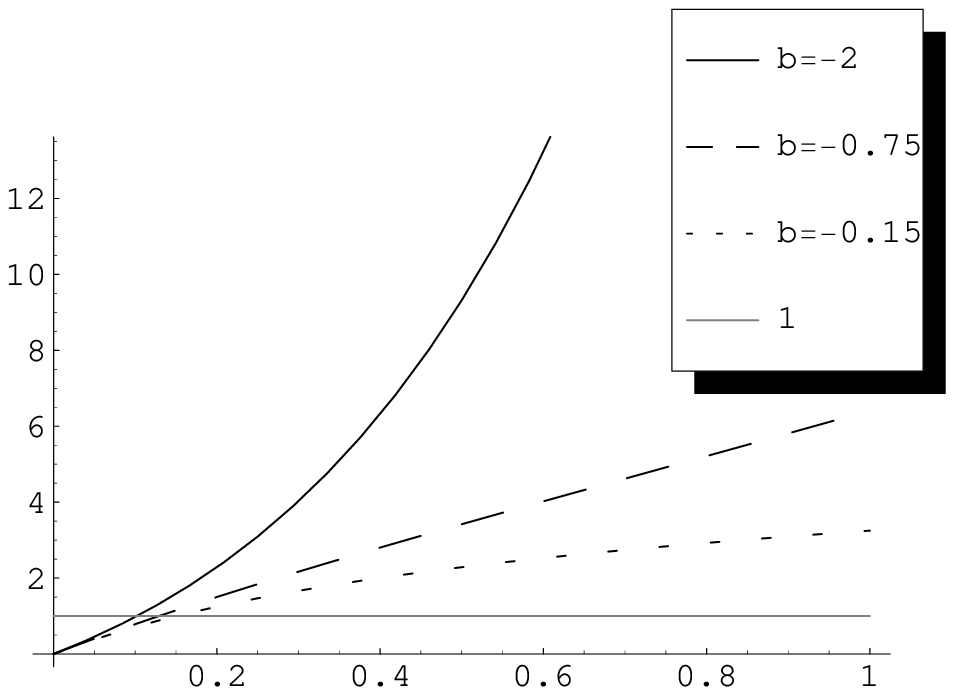}}

\caption{Left Figure: the function $NI(N)$ against the constant 1
when $a(N)$ is linear. For $b=0.5$ we have considered
$a(N)=0.5+N/8$, for $b=1.2$: $a(N)=0.4+N/100$ and for $b=8$:
$a(N)=6+N/100$. Right Figure: the function $NI(N)$ against the
constant 1 with $b<0$ and $a(N)=1+N$. Here $V_R = 1$; $V_F = 2$.}
\label{fig:stst2}
\end{figure}

\subsection{Case of $a(N)=a_0+a_1N$}
We now treat the case of
$a(N)=a_0+a_1 N$, with $a_0,a_1>0$ with $b>0$. Proceeding as above
we can obtain from \fer{eq:alteri} the expression of its
derivative
\begin{equation}
\label{eq:derivatei} I'(N)= - \f{d}{dN} \left[\f{V_0(N)}{
\sqrt{a(N)}}\right]\; \big(I_1(N)-I_2(N)\big) +
\f{d}{dN}\left(\f{1}{\sqrt{a(N)}} \right) \big( V_F\, I_1(N)-V_R
I_2(N) \big),
\end{equation}
where
$$
I_1(N)= \int_0^{\infty} e^{- s^2/2}  e^{ s\, w_F} ds \quad
\mbox{and} \quad I_2(N)= \int_0^{\infty} e^{- s^2/2} e^{s\, w_R}
ds\, .
$$
Therefore $I(N)$ is decreasing since
$$
\f{d}{dN} \left[\f{V_0(N)}{ \sqrt{a(N)}}\right] =\f{2\, b\,
a_0+b\, a_1\, N}{2\, (a_0+a_1\, N)^{3/2}}
>0
\qquad \mbox{and} \qquad \f{d}{dN}\left(\f{1}{\sqrt{a(N)}}\right)
=-\f{a_1}{(a_0+a_1\, N)^{3/2}} <0\, .
$$
Moreover, we can check that the computation of the limit
\eqref{eq:limnin} still holds. Actually, we have
\begin{align*}
\lim_{N\to \infty} \frac{N\, I(N)}{V_F-V_R} &= \lim_{N\to \infty}
\f{N}{\sqrt{a}} \int_0^\infty e^{-s^2/2}\, e^{-s\, b\, N/\sqrt{a}}
\ ds = \lim_{N\to \infty} \sqrt{\pi} \f{erfc\left(\f{b\,
N}{\sqrt{2\, a}}\right)}{ \f{e^{-\f{b^2\, N^2}{2\,
a}}}{\f{N}{\sqrt{2\, a}}}}
\\ &= \lim_{\alpha\to \infty} \sqrt{\pi}\f{erfc(b\, \alpha)}
{\f{e^{-b^2\, \alpha^2}}{\alpha}}=
 \lim_{\alpha\to \infty}
\sqrt{\pi} \f{-\f{2}{\sqrt{\pi}}\, b\, e^{-b^2\, \alpha^2}
}{\f{-2\, b^2 \alpha^2 e^{-b^2\, \alpha^2}-e^{-b^2\,
\alpha^2}}{\alpha^2}} =\f{1}{b}\, ,
\end{align*}
where we have used the change $\alpha=\f{N}{\sqrt{2\, a}}$ and
L'H\^opital's rule. In the case $b<0$, we can observe again by the
same proof as before that $I(N)\to\infty$ when $N\to \infty$, and
thus, by continuity there is at least one solution to $NI(N)=1$.
Nevertheless, it seems difficult to clarify perfectly the number
of solutions due to the competing monotone functions in
\eqref{eq:derivatei}.

The generalization of part of Theorem \ref{th:stationary} is
contained in the following result. We will skip its proof since it
essentially follows the same steps as before with the new
ingredients just mentioned.
\begin{corollary} Assume $h(v,N)=b\,N-v$, $a(N)=a_0+a_1 N$ with
$a_0,a_1>0$.
\begin{itemize}
\item[i)]  Under either the condition $b < V_F-V_R$, or the
conditions $b>0$ and $2 a_0 b + 2 a_1 V_R < (V_F-V_R)^2 V_R$, then
there exists at least one steady state solution to
\eqref{eq:nif1}-\eqref{eq:nif3}.

\item[ii)] If both $2 a_0 b + 2 a_1 V_R < (V_F-V_R)^2 V_R$ and $b
> V_F-V_R$ hold, then there are at least two steady states to
\eqref{eq:nif1}-\eqref{eq:nif3}.

\item[iii)] There is no steady state to
\eqref{eq:nif1}-\eqref{eq:nif3} for $b>
\max \big(2(V_F-V_R),2V_F\,I(0) \big)$.
\end{itemize}
\label{cor:stationary}
\end{corollary}

These behaviours are depicted in Figure \ref{fig:stst2}. Let us
point out that if $a$ is linear and $b<0$, $I(N)$ may have a
minimum for $N>0$.


\section{Linear case and relaxation}
\label{sec:linear}

We study specifically the linear case, $b=0$ and $a(N)=a$, i.e.,
\begin{equation}\label{eq:linnif}
\left\{ \begin{array}{l} \dis\frac{\partial p(v,t)}{\partial t}-
\frac{\partial}{\partial v} [v p(v,t)] - a_0 \frac{\p^2}{\p  v^2}
p(v,t) =  \delta(v-V_R) N(t), \qquad v\leq V_F,
\\[3mm]
p(V_F,t)=0, \qquad N(t):= -a_0 \frac{\p}{\p v} p(V_F,t) \geq 0,
\quad a_0>0,
\\[2mm]
\dis p(v,0)=p^0(v)\geq 0, \qquad \int_{-\infty}^{V_F} p^0(v) dv
=1.
\end{array} \right.
\end{equation}
For later purposes, we remind that the steady state $p_\infty(v)$
given in \eqref{stationaryplinear} satisfies
\begin{equation}
\label{eq:linstst} \left\{ \begin{array}{l} \dis
-\frac{\partial}{\partial v} [v p_\infty(v)] - a_0 \frac{\p^2}{\p
v^2}p_\infty(v) =  \delta(v-V_R) N_\infty, \qquad v\leq V_F,
\\[3mm]
p_\infty(V_F)=0, \qquad N_\infty:= -a_0 \frac{\p}{\p v}
p_\infty(V_F) \geq 0,
\\[2mm]
\dis \int_{-\infty}^{V_F} p_\infty(v) dv =1.
\end{array} \right.
\end{equation}
We will assume in this section that solutions of the linear
problem exist with the regularity needed in each result below and
such that for all $T>0$ there exists $C_T>0$ such that $p(v,t)\leq
C_T p_\infty(v)$ for all $0\leq t\leq T$. These solutions might be
obtained by the method developed in \cite{GG09} and will be
analysed elsewhere.

We prove that the solutions converge in large times to the unique
steady state $p_\infty(v)$. Two relaxation processes are involved
in this effect: dissipation by the diffusion term and dissipation
by the firing term. This is stated in the following result about
relative entropies for this problem.

\begin{theorem} \label{th:relentr}
Fast-decaying solutions to equation \eqref{eq:linnif} satisfy, for
any smooth convex function $G:\R^+\longrightarrow \R$, the
inequality
\begin{align}\label{eq:linnifentropy}
\frac{d }{d t}\int_{-\infty}^{V_F} p_\infty(v)
G\left( \f{p(v,t)}{ p_\infty(v)} \right) =\, & - N_\infty  \left[
G\left( \f{N(t)}{ N_\infty} \right) -ÊG\left( \f{p(v,t)}{
p_\infty(v)} \right) -\left( \f{N(t)}{ N_\infty} - \f{p(v,t)}{
p_\infty(v)}\right)G'\left( \f{p(v,t)}{ p_\infty(v)}
\right)\right]{
 |_{V_R}}
\nonumber\\[3mm]
& -a_0 \int_{-\infty}^{V_F} p_\infty(v) \; G''\left( \f{p(v,t)}{
p_\infty(v)} \right) \; \left[ \f{\p}{\p v} \left(\f{ p(v,t)}{
p_\infty(v)} \right)\right]^2\,dv \leq 0.
\end{align}
\end{theorem}

The following result is in fact standard on Poincar\'e
inequalities on $\R$ once $q$ and $p_\infty$ have been extended to
the full line by odd symmetry with respect to $V_F$ because
$p_\infty$ has a Gaussian behaviour at infinity thanks to
\eqref{stationaryplinear}, see \cite{ledoux}.

\begin{proposition} \label{th:poincare} There exists $\nu >0$ such that
\begin{equation*}
\nu \int_{-\infty}^{V_F} p_\infty(v) \left( \f{q(v)}{ p_\infty(v)}
\right)^2 \leq \int_{-\infty}^{V_F} p_\infty(v)  \; \left[
\f{\p}{\p v} \left(\f{ q(v)}{ p_\infty(v)} \right)\right]^2
\end{equation*}
for all functions $q$ such $\frac{q}{p_\infty}\in
H^1\big(p_\infty(v) dv\big)$.
\end{proposition}

Note that performing the even symmetry of $q$ with respect to
$V_F$ ensures that the extended function $\tilde q$ satisfies
$$
\int_{\R} \tilde q(v) \,dv =0 .
$$
These two theorems have direct consequences as in
\cite{RefMMP}.

\begin{corollary}[Exponential decay] Fast-decaying solutions to the equation
\eqref{eq:linnif} satisfy
$$
\int_{-\infty}^{V_F} p_\infty(v) \left( \f{p(v,t)-p_\infty(v)}{
p_\infty(v)} \right)^2 \leq e^{-2a_0\nu t} \int_{-\infty}^{V_F}
p_\infty(v) \left( \f{p^0(v)-p_\infty(v)}{ p_\infty(v)} \right)^2.
$$
\end{corollary}

\proof Taking $q=p(v,t)-p_\infty(v)$ and $G(x)=(x-1)^2$ in the
relative entropy inequality \fer{eq:linnifentropy}, we obtain
\begin{equation*}
\frac{d }{d t}\int_{-\infty}^{V_F}
p_\infty(v) \left( \f{p(v,t)}{ p_\infty(v)} -1\right)^2
\le
 -2a_0 \int_{-\infty}^{V_F}  p_\infty(v) \;
 \left[ \f{\p}{\p v} \left(\f{ p(v,t)}{ p_\infty(v)}-1 \right)\right]^2 .
\end{equation*}
Poincar\'e's  inequality in Proposition \ref{th:poincare} bounds
the right hand side on the previous inequality
\begin{equation*}
\frac{d }{d t}\int_{-\infty}^{V_F}
p_\infty(v) \left( \f{p(v,t)}{ p_\infty(v)} -1\right)^2
\le
 -2a_0 \mu\int_{-\infty}^{V_F}  p_\infty(v) \;
  \left(\f{ p(v,t)}{ p_\infty(v)}-1 \right)^2
 .
\end{equation*}
Finally,  the Gronwall lemma directly gives the result.
\qed

\

The proof of Theorem \ref{th:relentr} is based on the following
computations.

\begin{lemma}\label{lin-EDPs}
Given $p$ a fast-decaying solution of \eqref{eq:linnif},
$p_\infty$ given by \eqref{stationaryplinear} and $G(\cdot)$ a
convex function, then the following relations hold:
\begin{equation}\label{eq:linppinf}
\frac{\partial }{\partial t} \f{p}{ p_\infty} - \left( v+ \f{2
a_0}{p_\infty} \frac{\partial }{\partial v} p_\infty \right)
\frac{\partial }{\partial v} \f{p}{ p_\infty} -a_0
\frac{\partial^2 }{\partial v^2} \f{p}{ p_\infty}= \f{N_\infty}{
p_\infty} \; \delta \big( v-V_R\big) \left( \frac{N }{N_\infty} -
\f{p}{ p_\infty} \right),
\end{equation}
\begin{align}
\label{eq:linG} \frac{\partial }{\partial t} G\left( \f{p}{
p_\infty} \right)&- \left( v+ \f{2 a_0}{p_\infty} \frac{\partial
}{\partial v} p_\infty \right) \frac{\partial }{\partial v}
G\left( \f{p}{ p_\infty} \right) -a_0 \frac{\partial^2 }{\partial
v^2}G\left( \f{p}{ p_\infty} \right)
\\
&=-a_0 G''\left( \f{p}{ p_\infty} \right)  \; \left( \f{\p}{\p v}
\f{ p}{ p_\infty} \right)^2  +  \f{N_\infty}{ p_\infty} \; \delta
\big( v-V_R\big) \left( \frac{N }{N_\infty} - \f{p}{ p_\infty}
\right)
 G'\left( \f{p}{ p_\infty}\right), \nonumber
\end{align}
\begin{align}\label{eq:linGpinf}
\frac{\partial }{\partial t} p_\infty & G\left( \f{p}{ p_\infty}
\right) - \frac{\partial }{\partial v} \left[ vp_\infty G\left(
\f{p}{ p_\infty} \right)
 \right] -a_0 \frac{\partial^2 }{\partial v^2}
\left[  p_\infty G\left( \f{p}{ p_\infty} \right) \right]
\\
&=-a_0 p_\infty G''\left( \f{p}{ p_\infty} \right)  \; \left(
\f{\p}{\p v} \f{ p}{ p_\infty} \right)^2  + N_\infty\; \delta
\big( v-V_R\big) \left[ \left( \frac{N }{N_\infty} - \f{p}{
p_\infty} \right)
 G'\left( \f{p}{ p_\infty}\right)  +G\left( \f{p}{ p_\infty} \right)
 \right].\nonumber
\end{align}
\end{lemma}

\proof
Since $\f{\p}{\p v}\left(\f{p}{p_\infty} \right)=
\f{1}{p_\infty}\f{\p \, p}{\p v}-\f{p}{p_\infty^2}\f{\p  p_\infty}{\p v}$ we obtain
$$
\f{\p p}{\p v} =p_\infty \f{\p}{\p v}\left(\f{p}{p_\infty}\right)+
\f{p}{p_\infty}\f{\p p_\infty}{\p v}.
$$
and
$$
\f{\p^2 p}{\p v^2}=
p_\infty \f{\p^2}{\p v^2}\left(\f{p}{p_\infty}\right)+
2\f{\p }{\p v}\left(\f{p}{p_\infty}\right)
\f{\p p_\infty}{\p v}
+
\f{p}{p_\infty}\f{\p^2 p_\infty}{\p v^2}.
$$
Using these two expressions in
$$
\f{\p}{\p t}\left(\f{p}{p_\infty} \right)=
\f{1}{p_\infty}\f{\p p}{\p t}=
\f{1}{p_\infty}\left\{
 \delta(v=V_R) N(t)
+ \frac{\partial}{\partial v} [v p(v,t)] + a_0 \frac{\p^2}{\p
v^2} p(v,t) \right\}
$$
we obtain (\ref{eq:linppinf}).

Equation (\ref{eq:linG}) is a consequence of Equation (\ref{eq:linppinf})
and the following expressions for the partial derivatives of
$G\left(\f{p}{p_\infty} \right)$:
$$
\f{\p}{\p t} G\left(\f{p}{p_\infty} \right)=
G'\left(\f{p}{p_\infty} \right)\f{\p}{\p t}\left(\f{p}{p_\infty} \right),
\
\f{\p}{\p v} G\left(\f{p}{p_\infty} \right)=
G'\left(\f{p}{p_\infty} \right)\f{\p}{\p v}\left(\f{p}{p_\infty} \right)
$$
and
$$
\f{\p^2}{\p v^2} G\left(\f{p}{p_\infty} \right)=
G''\left(\f{p}{p_\infty} \right)\left(\f{\p}{\p v}\left(\f{p}{p_\infty}
\right)\right)^2+
G'\left(\f{p}{p_\infty} \right)\f{\p^2}{\p v^2}\left(\f{p}{p_\infty} \right).
$$
Finally,  Equation (\ref{eq:linGpinf}) is obtained using Equation
(\ref{eq:linG}) and the fact that $p_\infty$ is solution of (\ref{eq:linstst}).
\qed
\bigskip

\noindent {\bf Proof Theorem \ref{th:relentr}.} We  integrate from
$-\infty$ to $V_F-\alpha$ in (\ref{eq:linGpinf}) and let $\alpha$
tend to $0^+$ and use L'H\^opital's rule
\begin{equation}\label{limbc}
 \lim_{v\to V_F} \f{ p(v,t)}{ p_\infty(v)} =  \lim_{v\to V_F} \f{ \frac{\partial p}{\partial v} (v,t)}{ \frac{\partial p_\infty}{\partial v}  (v)}= \f{N(t)}{N_\infty}.
\end{equation}
Since $p(v,t) \leq C_T p_\infty$ with $0\leq t\leq T$, then
\begin{align*}
\frac{d }{d t} \int_{-\infty}^{V_F} & p_\infty G\left( \f{p}{
p_\infty} \right) \,dv-a_0 \frac{\partial }{\partial v} \left[
p_\infty G\left( \f{p}{ p_\infty} \right) \right] |_{V_F}
\\
&=-a_0 \int_{-\infty}^{V_F} p_\infty G''\left( \f{p}{ p_\infty}
\right)  \; \left( \f{\p}{\p v} \f{ p}{ p_\infty} \right)^2\,dv  +
N_\infty \left[ \left( \frac{N }{N_\infty} - \f{p}{ p_\infty}
\right) G'\left( \f{p}{ p_\infty}\right)  +G\left( \f{p}{
p_\infty} \right) \right] |_{V_R}.
\end{align*}
The Dirichlet boundary condition \eqref{eq:nif2} implies that
$$
-a_0 \frac{\partial }{\partial v} \left[  p_\infty G\left( \f{p}{
p_\infty} \right) \right] |_{V_F} =-a_0  \frac{\partial
p_\infty}{\partial v}  \ G\left( \f{p}{ p_\infty} \right)
|_{V_F} =N_\infty G\left( \f{N(t)}{N_\infty} \right),
$$
where we used that
$$
p_\infty \frac{\partial}{\partial\, v}
G\left(\frac{p}{p_\infty}\right)|_{V_F} = p_\infty
G'\left(\frac{p}{p_\infty}\right) \frac{-Np_\infty+N_\infty\,
p}{p_\infty^2\,a_0} |_{V_F}=
 G'\left(\frac{q}{p_\infty}\right)
\left(\frac{-N}{a_0}+\frac{N_\infty}{a_0}
\frac{p}{p_\infty}\right)|_{V_F}=0,
$$
due to \eqref{limbc}. Collecting all terms leads to the desired
inequality. \qed

\section{Numerical results}
\label{sec:num}

We consider an explicit method to simulate the numerical
approximation for the NNLIF \eqref{eq:nif1}. We base our algorithm
on standard shock-capturing methods for the advection term
and second-order finite differences for the second-order term.
More precisely, the first order term is approximated by finite
difference WENO-schemes \cite{S}.

\begin{figure}[ht]
\epsfxsize=3.in \hbox to \hsize{\epsfbox{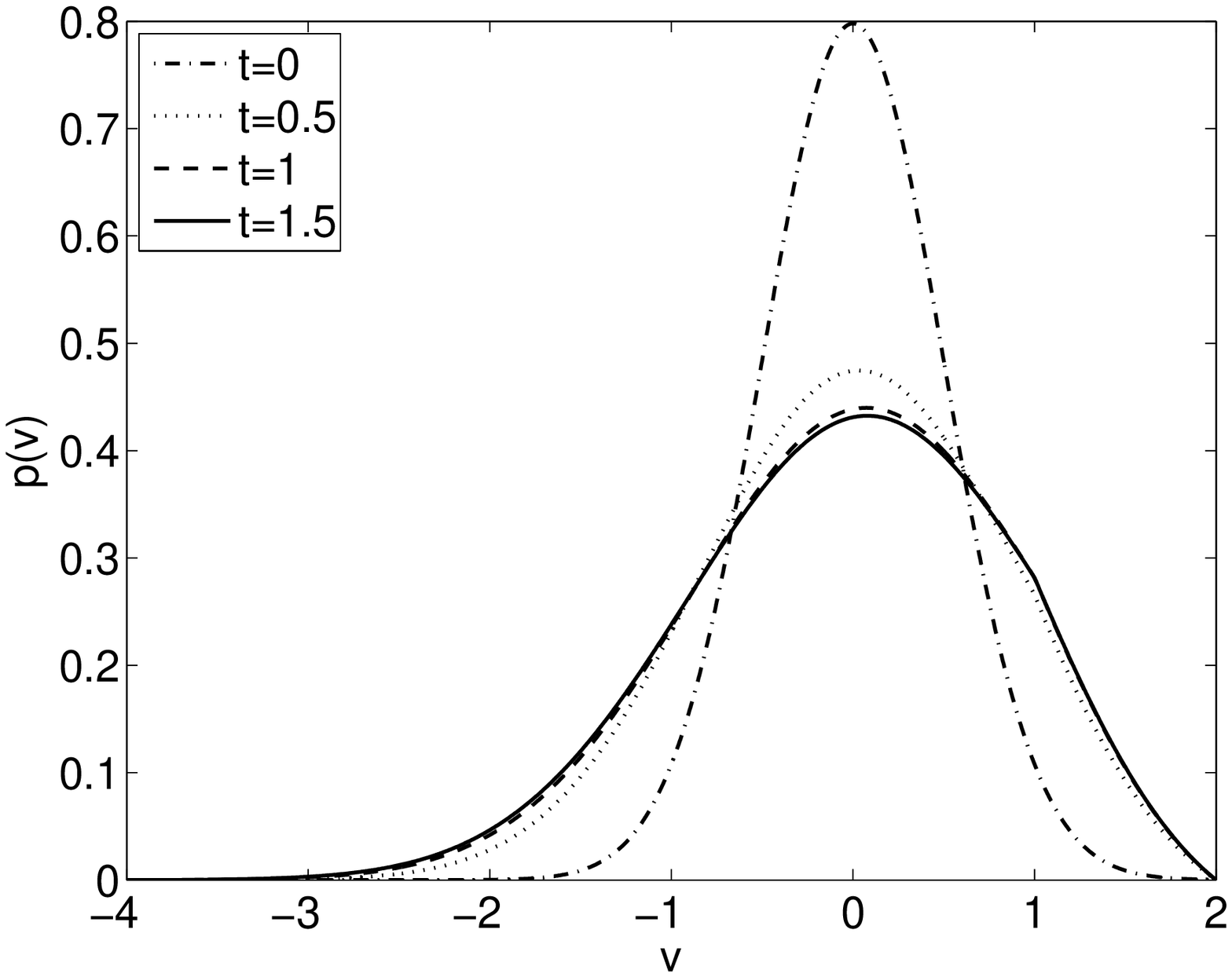}
\epsfxsize=3.in \hfil \epsfbox{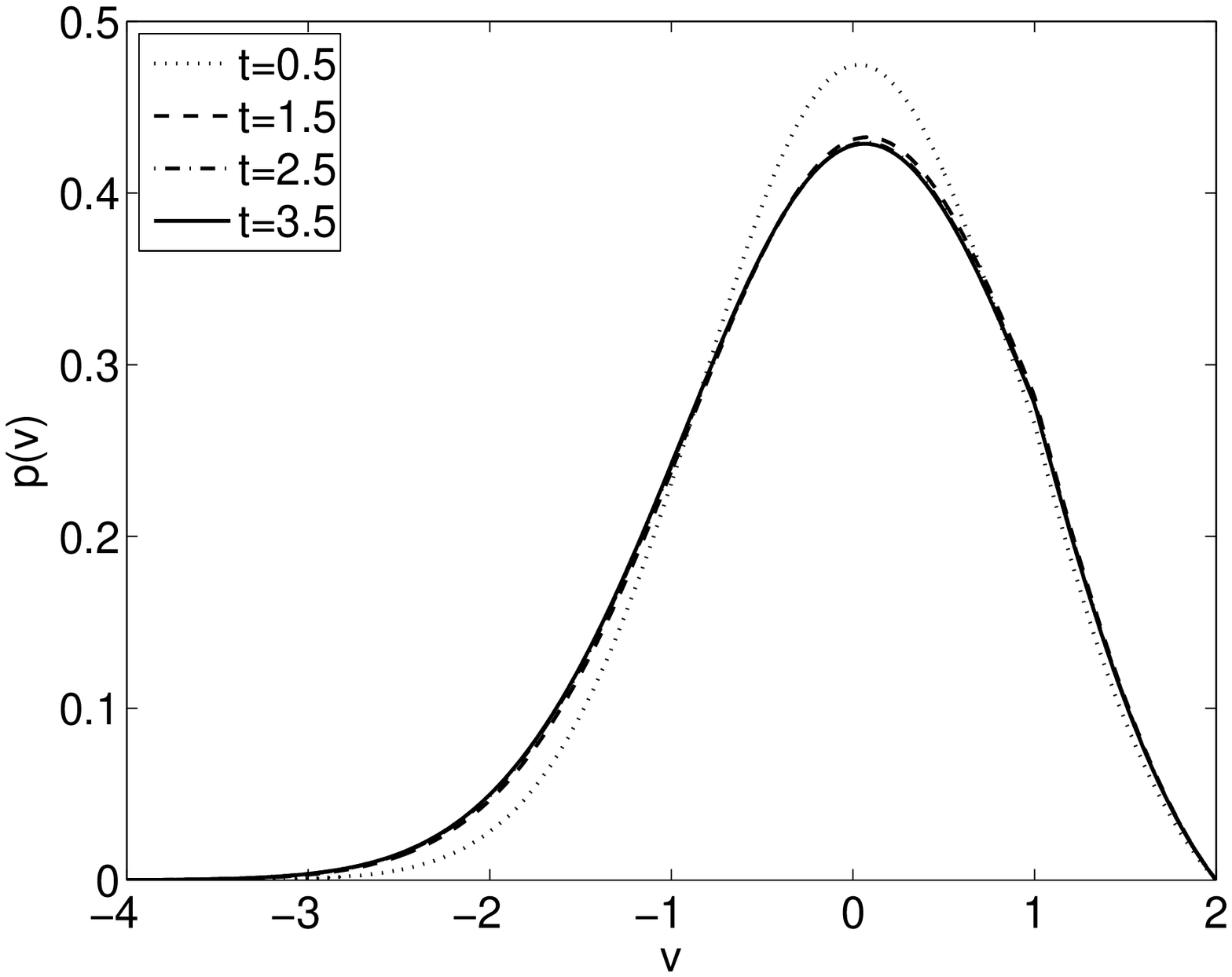}}
\vspace{-5mm} \caption{ Distribution functions $p(v,t)$ for
$b=0.5$ and $a=1$ at differents times.} \label{distr05}
\end{figure}

The time evolution is performed with a TVD Runge-Kutta scheme. Other
finite difference scheme for the Fokker-Planck equation has been
used such as the Chang-Cooper method \cite{BCD} similarly to the
approximation studied in \cite{CCTa} for a model with variable
voltage and conductance. The Chang-Cooper method presents
difficulties when the firing rate becomes large and the diffusion
coefficient $a(N)$ is constant. To discuss this, we have just to
remind the reader that the Chang-Cooper method performs a kind of
$\theta$-finite difference approximation of $p/M$ where $M$ is a
Maxwellian in the kernel of the linear Fokker-Planck operator.
Whenever $a(N)$ is constant, $b>0$ and $N$ is large, the drift of
the Maxwellian, in terms of which is rewritten the Fokker-Planck
equation, practically vanishes on the interval $(-\infty,V_F]$ and
this particular Chang-Cooper method is not suitable.

In our simulations we  consider a uniform mesh in $v$, for $v\in
[V_{min}, V_F]$. The value $V_{min}$ (less than $V_R$) is adjusted
in the numerical experiments to fulfill that $p(V_{min},t)\approx
0$, while $V_F$ is fixed to 2 and $V_R=1$. Most of our initial
data are Maxwellians:
$$
p_0(v)=
\frac{1}{\sigma_0\sqrt{2\pi}}e^{-\frac{(v-v_0)^2}{2\sigma_0^2}},
$$
where the mean $v_0$ and the variance $\sigma_0^2$ are chosen
according to the analyzed phenomenon. When the system has two
steady states, we also take as initial data the profiles given by
\eqref{stationaryp} with $N$ an approximate value of the
stationary firing rate, in order to start close to the stationary
state with larger firing rate.

\begin{figure}[ht]
\epsfxsize=3.in \hbox to \hsize{\epsfbox{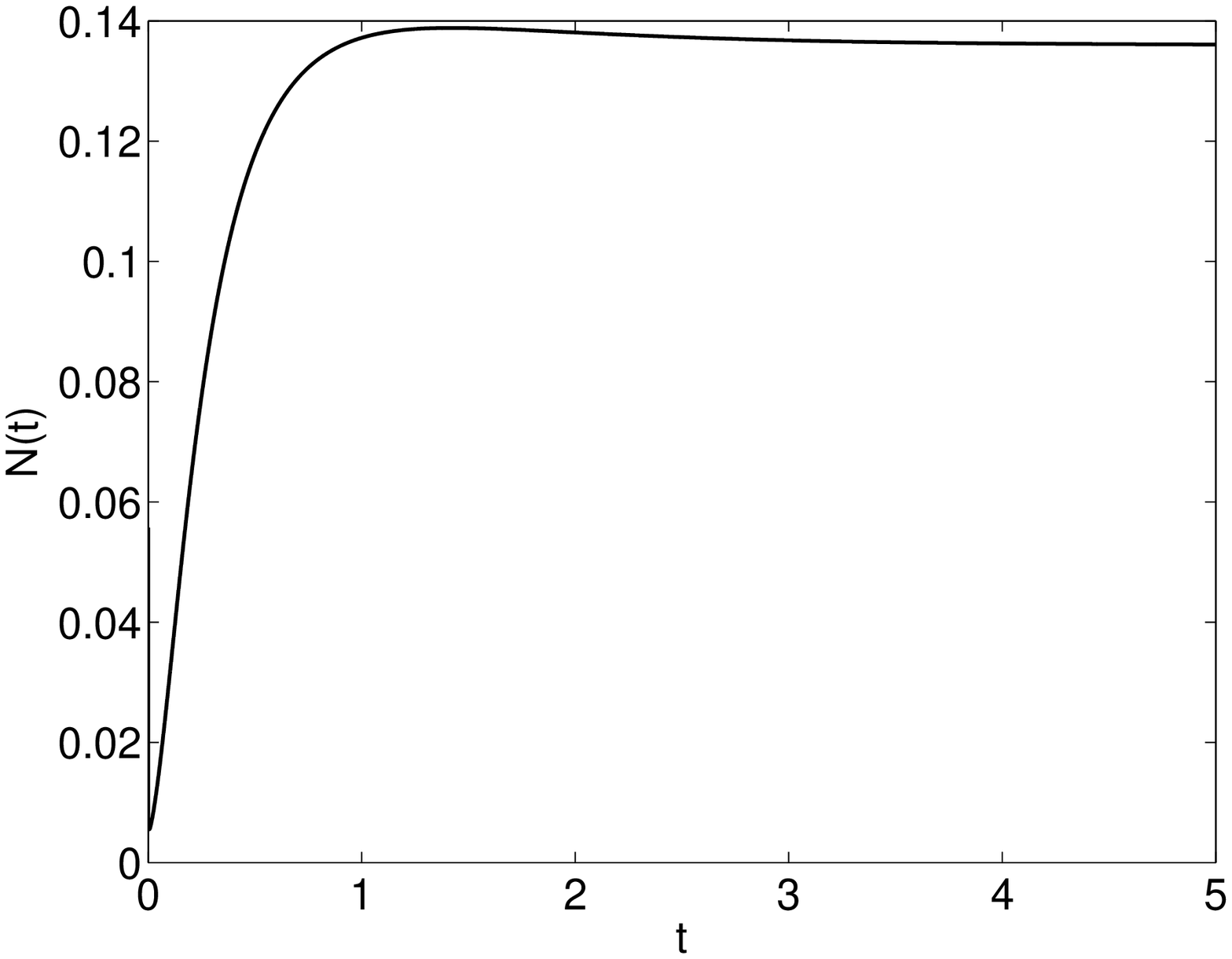}
\epsfxsize=3.in \hfil \epsfbox{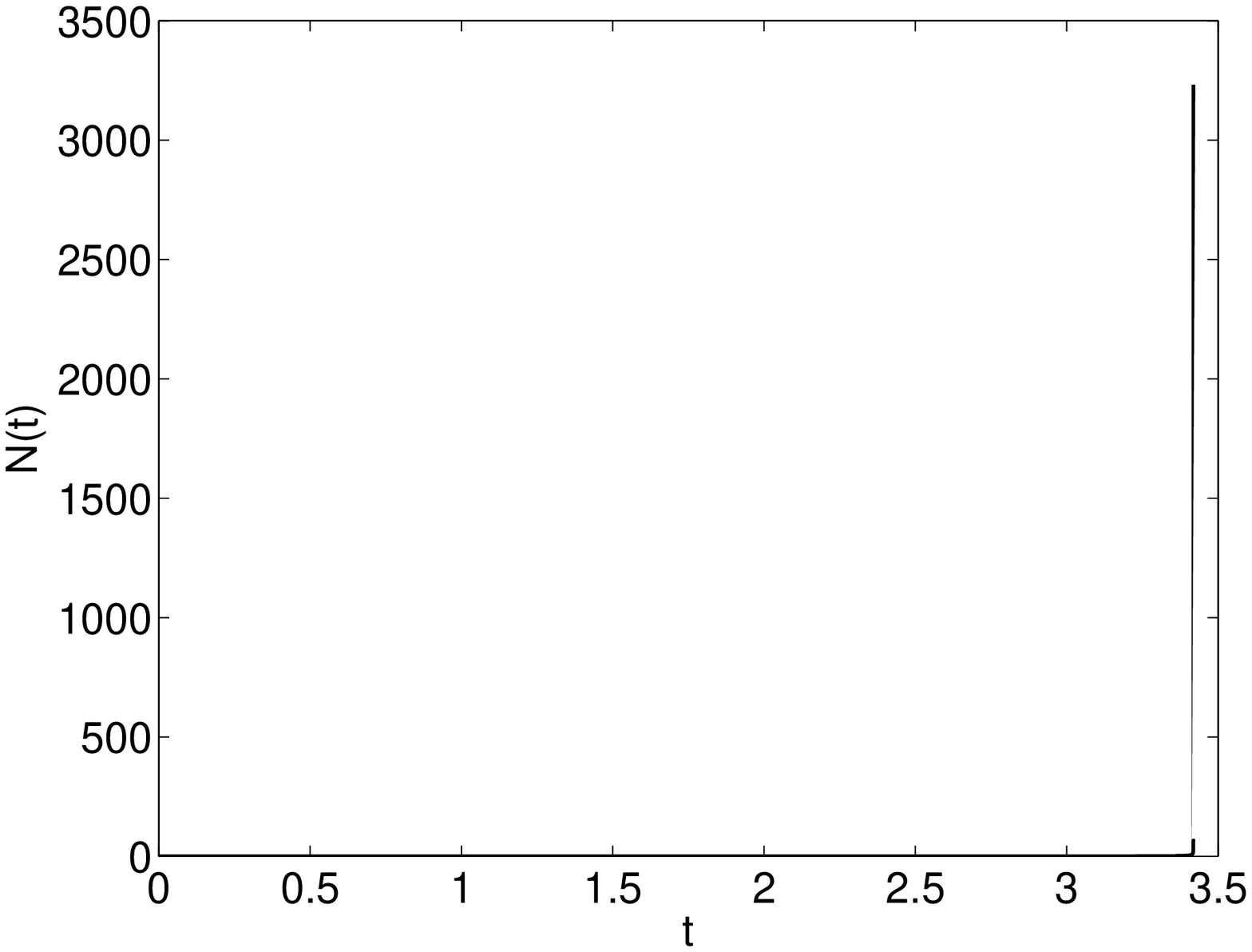}}

\epsfxsize=3.in \hbox to \hsize{\epsfbox{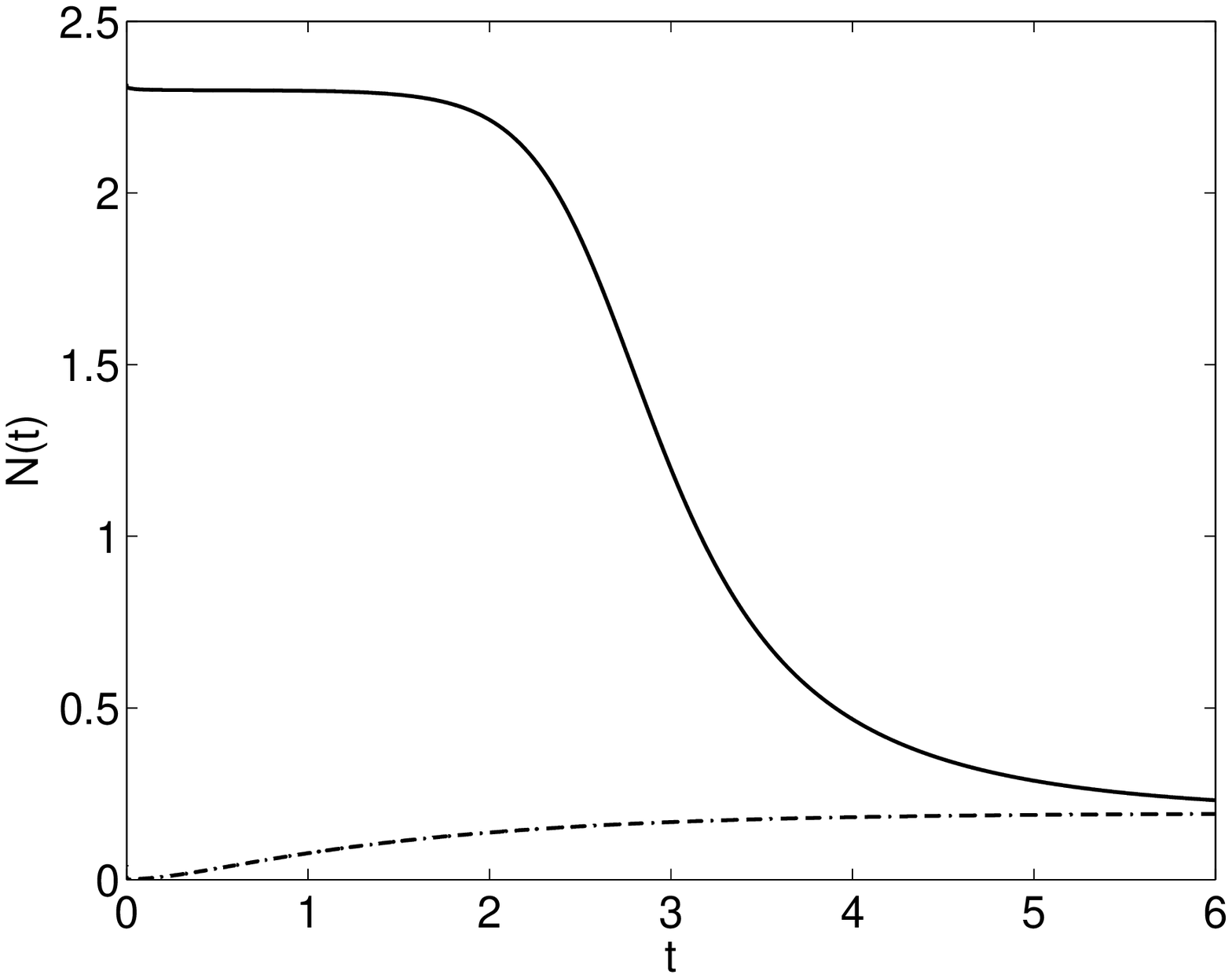}
\epsfxsize=3.in \hfil \epsfbox{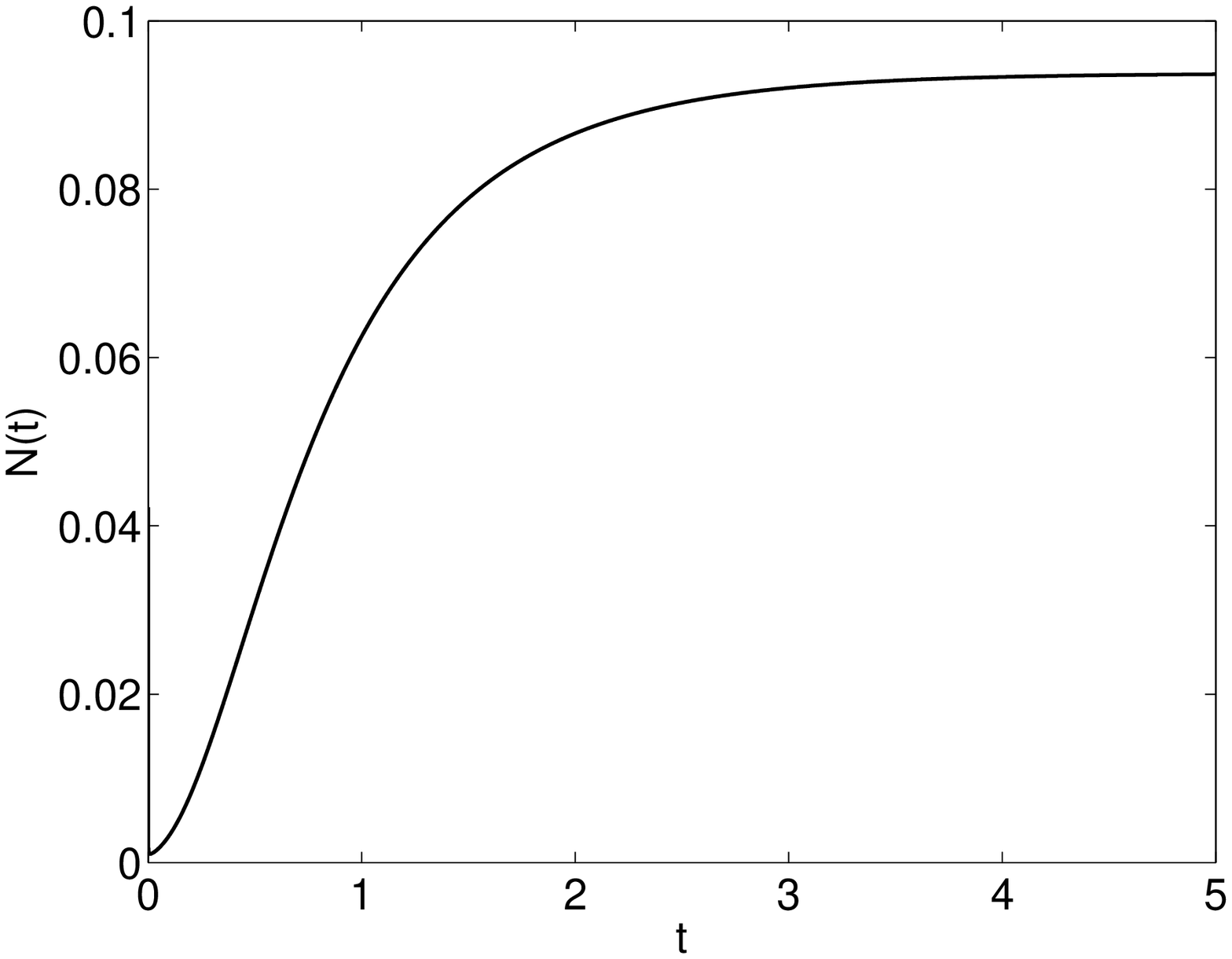}} \vspace{-5mm}
 \caption{Firing rates $N(t)$ for $a=1$.
Top left: $b=0.5$ with initial data a Maxwellian with: $v_0=0$ and
$\sigma_o^2=0.25$. Top right: $b=3$ with initial data a Maxwellian
with: $v_0=-1$ and $\sigma_o^2=0.5$. Bottom left: $b=1.5$
considering two different initial data: a Maxwellian with:
$v_0=-1$ and $\sigma_o^2=0.5$ and a  profile given by the
expression \eqref{stationaryp} with $N=2.31901$. Bottom right:
$b=-1.5$ with initial data a Maxwellian with: $v_0=-1$ and
$\sigma_o^2=0.5$. The top right case seems to depict a blow-up
phenomena demonstrated in Theorem \ref{th:bu}.}
\label{firing-aconst}
\end{figure}

{\em Steady states.-} As we show in Section \ref{sec:stst}, for
$b$ positive there is a range of values for which there are either
one or two or no steady states. With our simulations we can
observe all the cases represented in Figures \ref{fig:stst} and
\ref{fig:stst2}.

In Figure \ref{distr05} we show the time evolution of the
distribution function $p(v,t)$, in the case of $a=1$ and $b=0.5$, considering
as initial data a Maxwellian with $v_0=0$ and $\sigma_0^2=0.25$.
We observe that the solution after 3.5 time units numerically
achieves the steady state with the imposed tolerance. The top left
subplot in Figure \ref{firing-aconst} describes the time evolution
of the firing rate, which becomes constant after some time. This
clearly corresponds to the case of a unique locally asymptotically
stable stationary state. Let us remark that in the right subplot
of Figure \ref{distr05}, we can observe the Lipschitz behavior of
the function at $V_R$ as it should be from the jump in the flux
and thus on the derivative of the solutions and the stationary
states, see Section \ref{sec:stst}.

For $b=1.5$, we proved in Section \ref{sec:stst} that there are
two steady states. With our simulations we can conjecture that the
steady state with larger firing rate is unstable. However the
stationary solution with low firing rate is locally asymptotically
stable. We illustrate this situation in the bottom left subplot in
Figure \ref{firing-aconst}. Starting with a firing rate close to
the high stationary firing value, the solution tends to the low
firing stationary value.

\begin{figure}
\epsfxsize=3.in \hbox to \hsize{\epsfbox{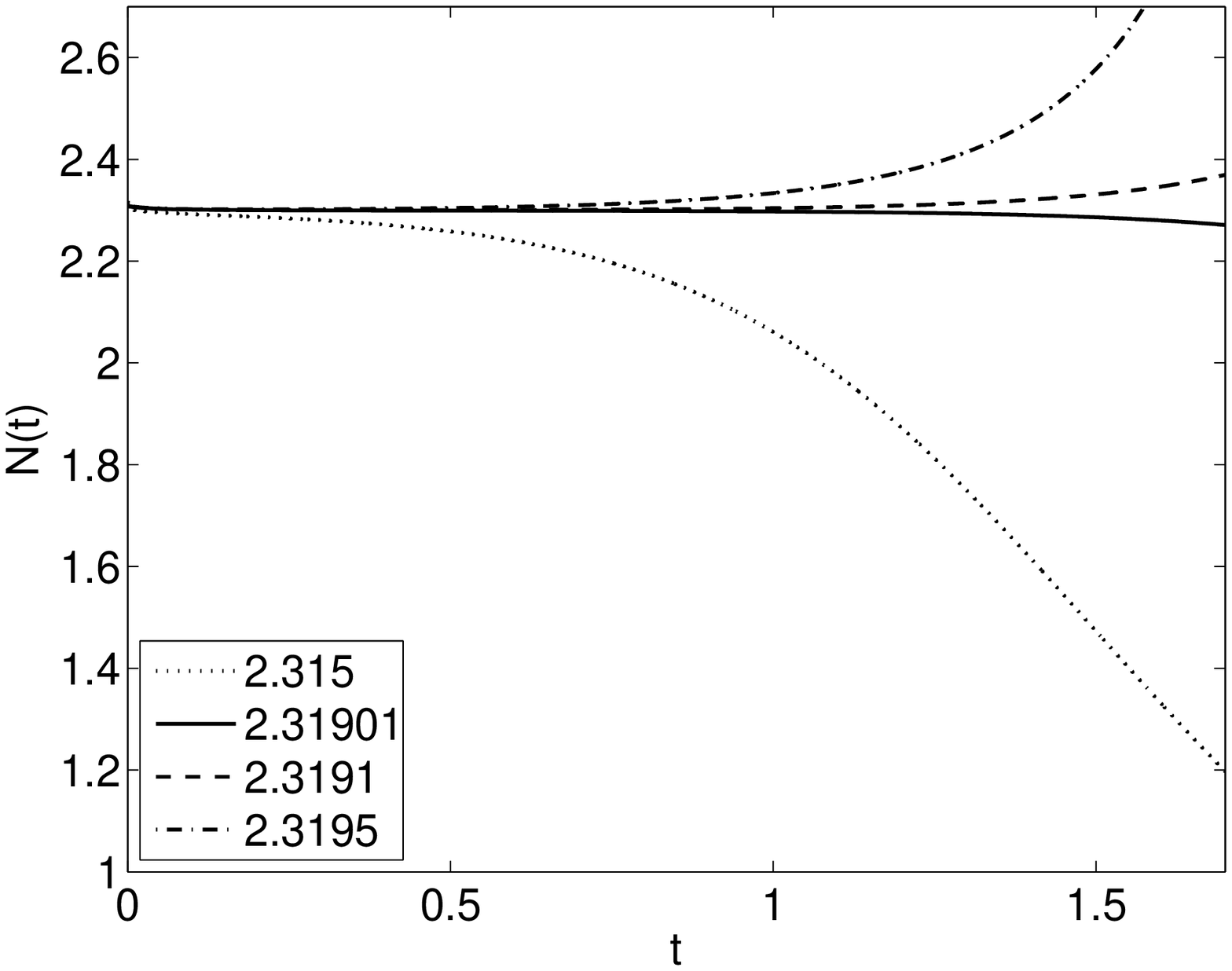}
\epsfxsize=3.in \hfil \epsfbox{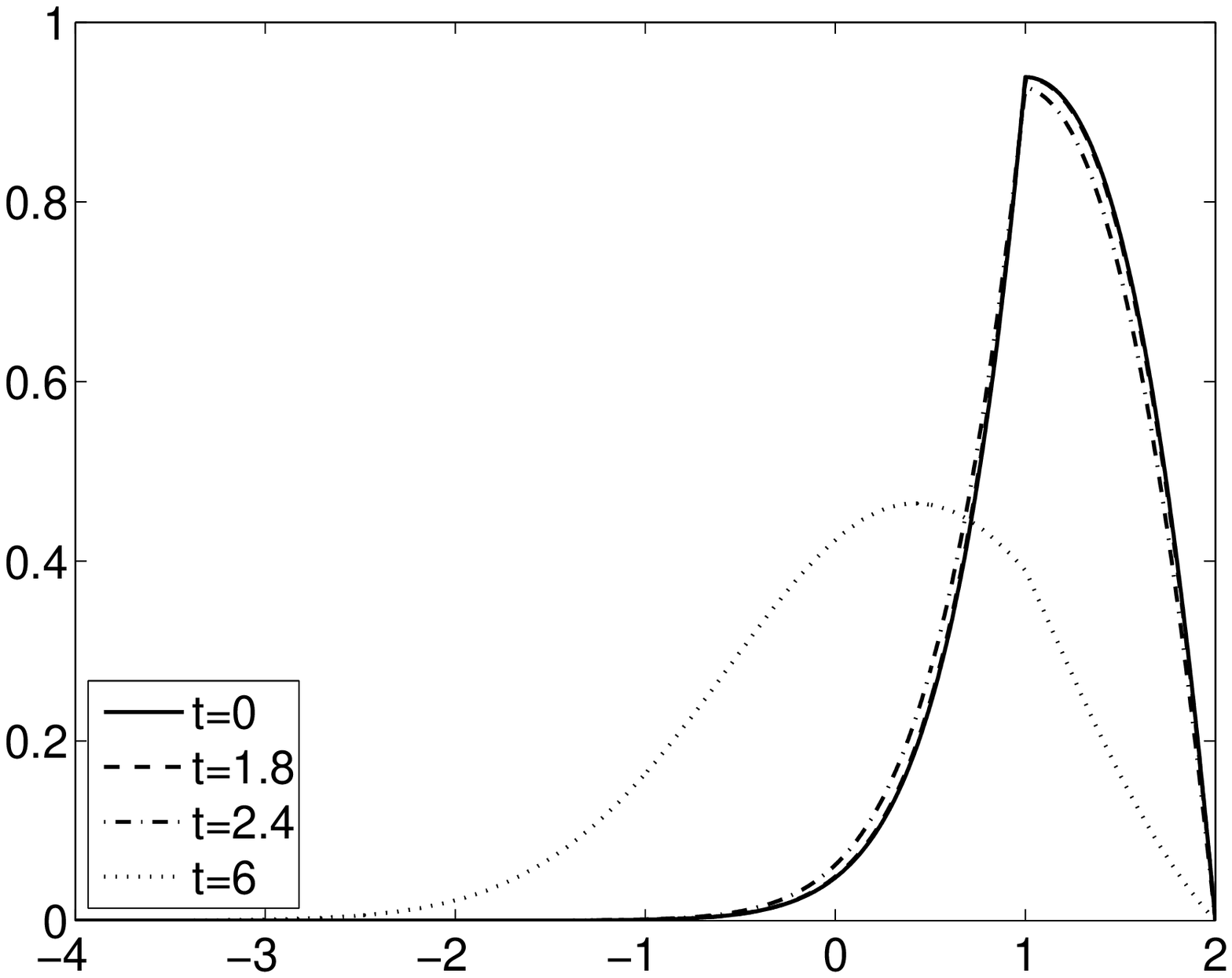}}

\caption{For $b=1.5$ and $a(N)=1$ figures show unstability
of the steady state with higher firing rate.
Left: Evolution on time of the firing rate considering different
initial firing rate.
Right: Evolution on time of the distribution function with initial
firing rate 2.31901. In both figures we have considered  $V_R = 1$; $V_F
= 2$.
} \label{fig:unstability}
\end{figure}

In Figure \ref{fig:unstability} we analyze in more details the
behavior of the steady state with larger firing rate. The left
subplot presents the evolution on time of the firing rate for
different distribution function starting with profiles given by
the expression \eqref{stationaryp} with $N$ an approximate value
of the stationary firing rate. We show that, depending of the
initial firing rate considered, its behavior is different: tends
to the lower steady state or goes to infinity. The firing rate for
the solution with initial $N_0=2.31901$ remains almost constant
for a period of time. Observe in Figure \ref{fig:unstability} that
the difference between the initial data and the distribution
function at time $t=1.8$ is almost negligible. However, the system
evolves slowly and at $t=6$ the distribution is very close to the
the lower steady state, see the bottom left subplot in Figure
\ref{firing-aconst}.

In the bottom right subplot of Figure \ref{firing-aconst} we
observe the evolution for a negative value of $b$, where we know
that there is always a unique steady state, and its local
asymptotic stability seems clear from the numerical experiments.

\

\begin{figure}[ht]
\epsfxsize=3.in \hbox to \hsize{\epsfbox{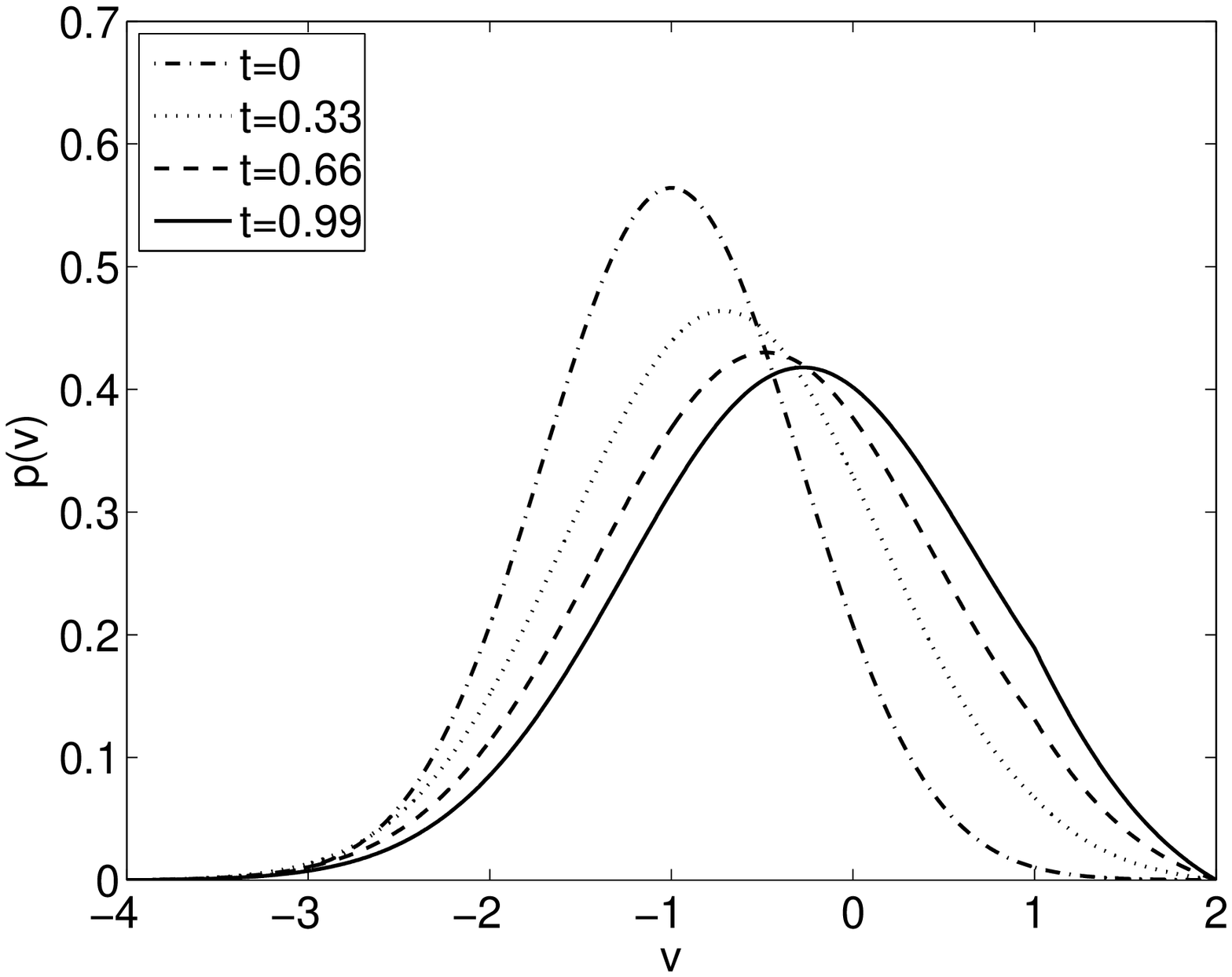}
\epsfxsize=3.in \hfil \epsfbox{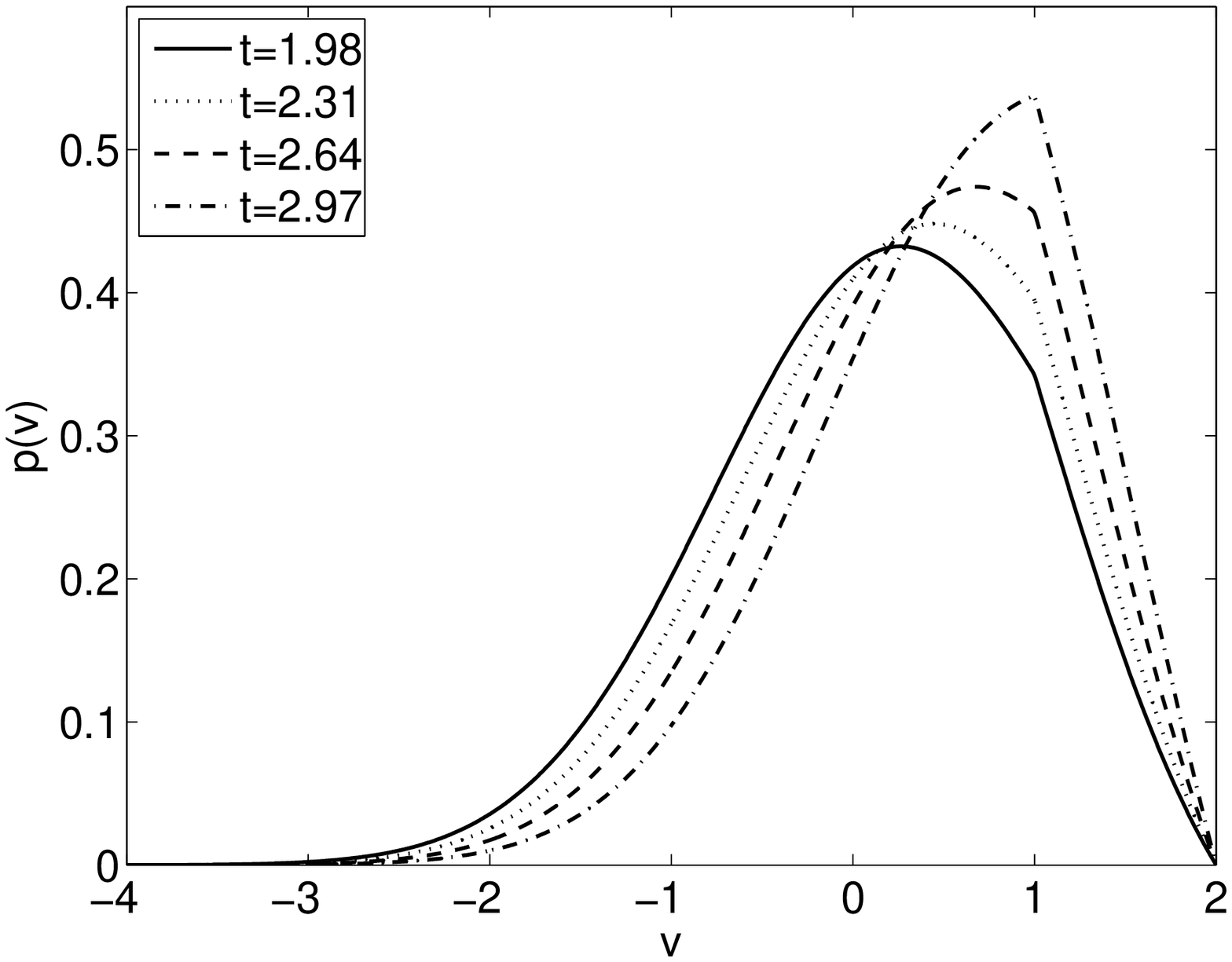}}
\vspace{-5mm}
\caption{Distribution functions $p(v,t)$ for  $a=1$  and $b=3$ at different
times. See Figure \ref{firing-aconst} for the corresponding plots of  $N(t)$.} \label{distr3}
\end{figure}

\

{\em No steady states.-} Our results in Section \ref{sec:stst}
indicate that there are no steady states for $b=3$. In Figure
\ref{distr3} we observe the evolution on time of the distribution
function $p$. In Figure \ref{firing-aconst} (right top) we show
the time evolution of the firing rate, which seems to blow up in
finite time. We observe how the distribution function becomes more
and more picked at $V_R$ and $V_F$ producing an increasing value
of the firing rate.

\

\begin{figure}[ht]
\epsfxsize=3.in \hbox to
\hsize{\epsfbox{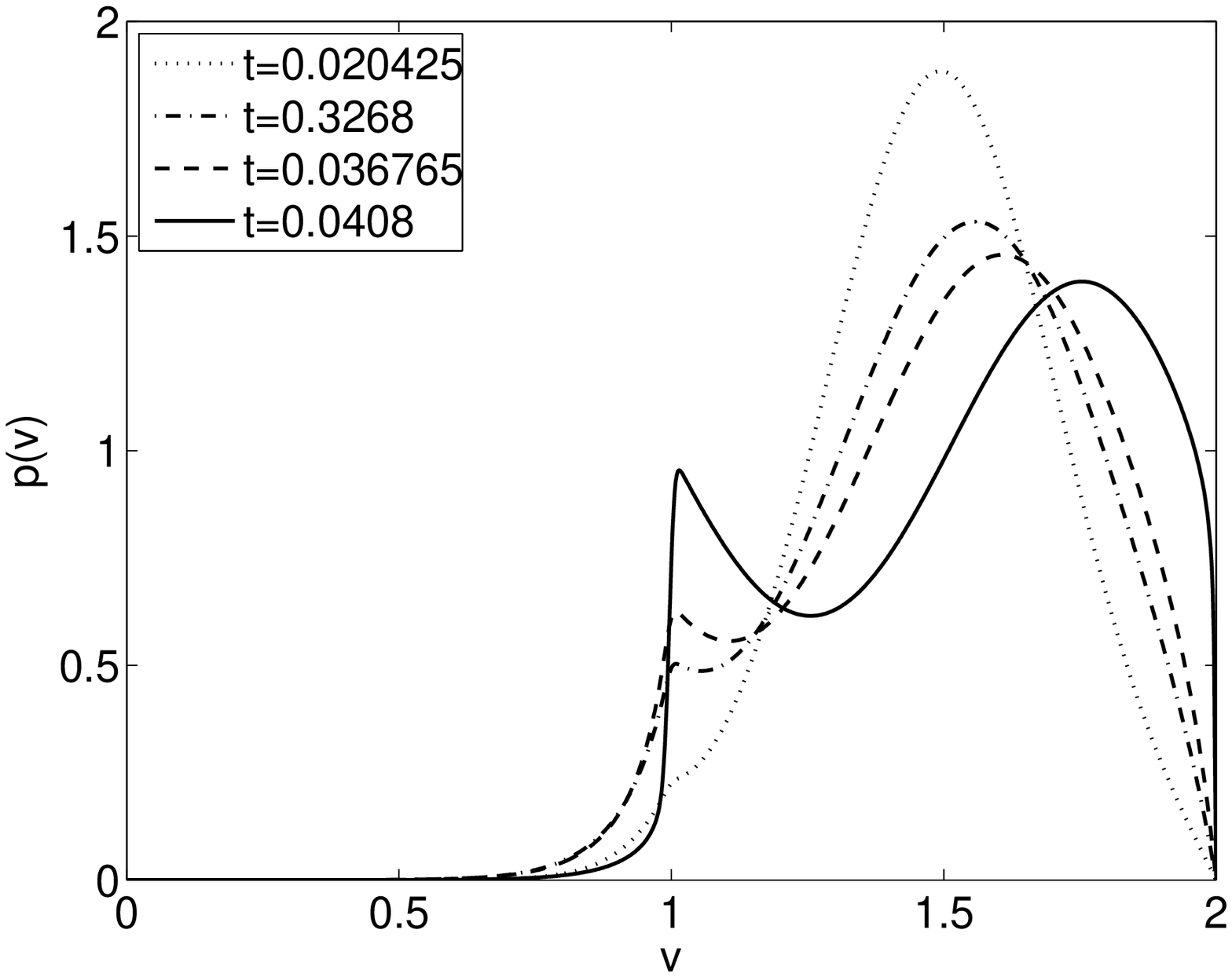} \epsfxsize=3.in
\hfil \epsfbox{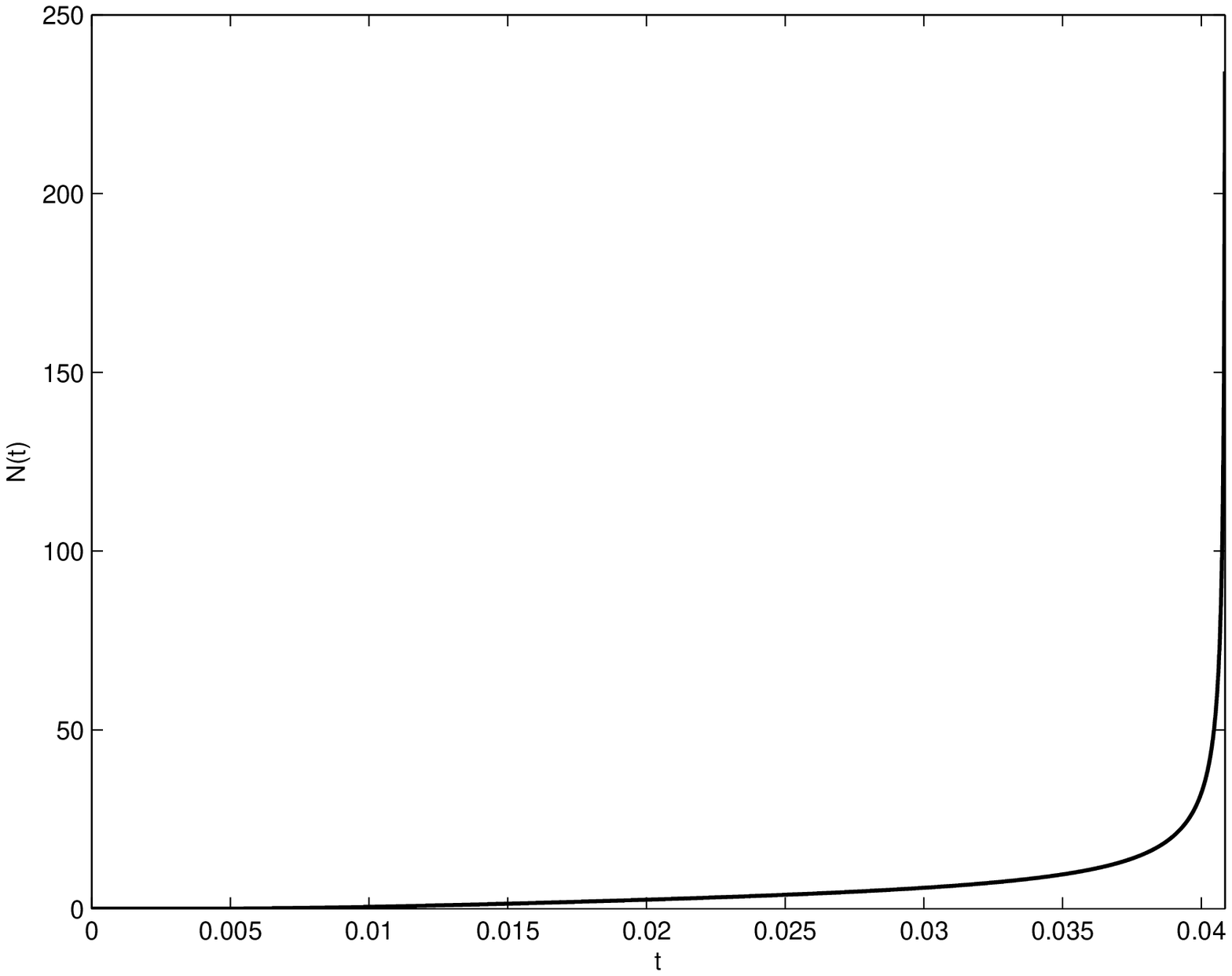}}
\vspace{-5mm}
\caption{Parameter values are  $a=1$ and $b=1.5$ and this corresponds to two steady states.
Left: Evolution of the distribution function $p(v,t)$ in time of
an initial Maxwellian centered at $v=1.5$ and with variance $0.005$.
Right: Time evolution of the firing rate; again we observe numerically a blow-up behaviour for an initial data enough concentrated near $V_F$.}
\label{blowup}\end{figure}

\begin{figure}[ht]
\epsfxsize=3.in \hbox to
\hsize{\epsfbox{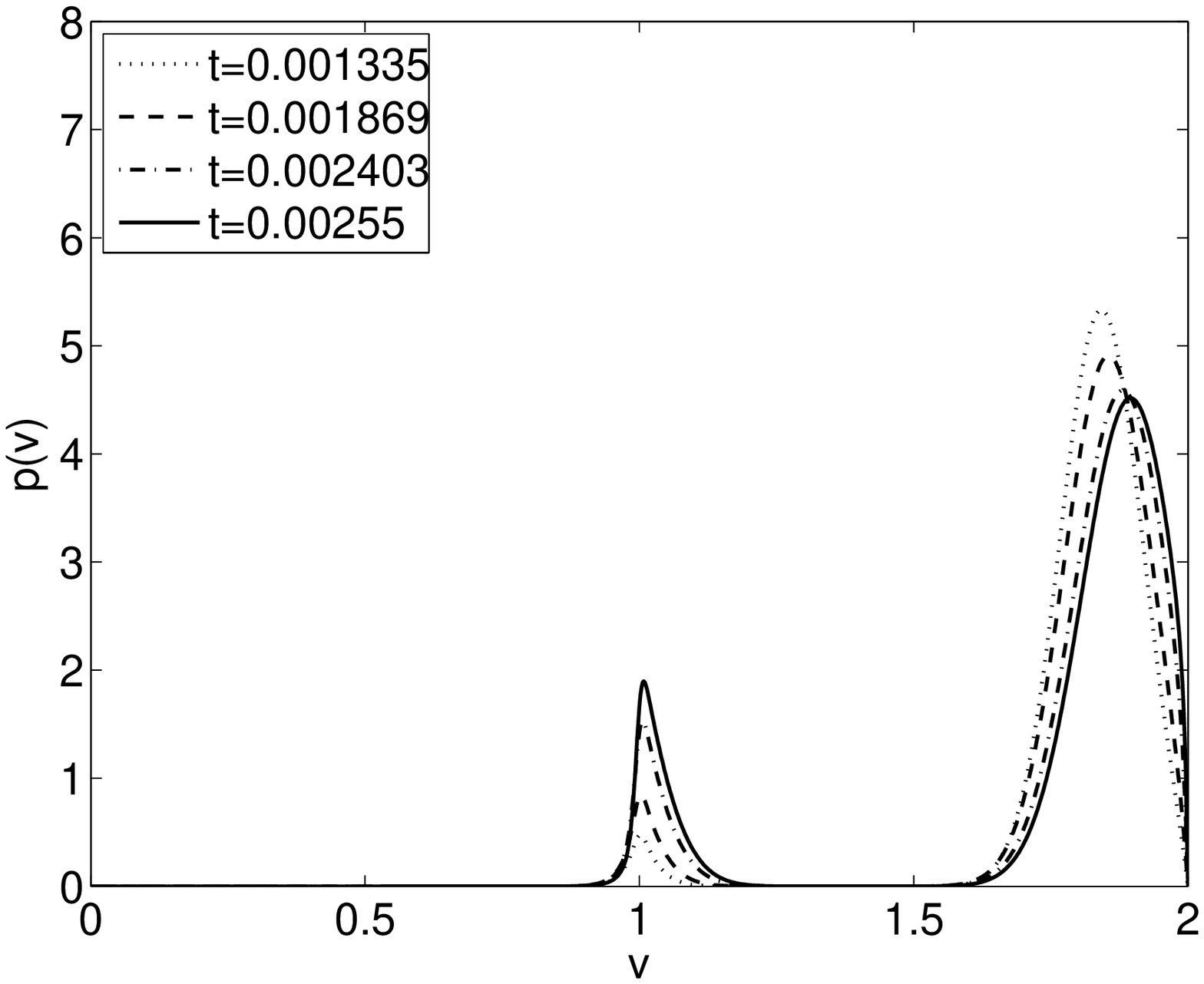} \epsfxsize=3.in
\hfil \epsfbox{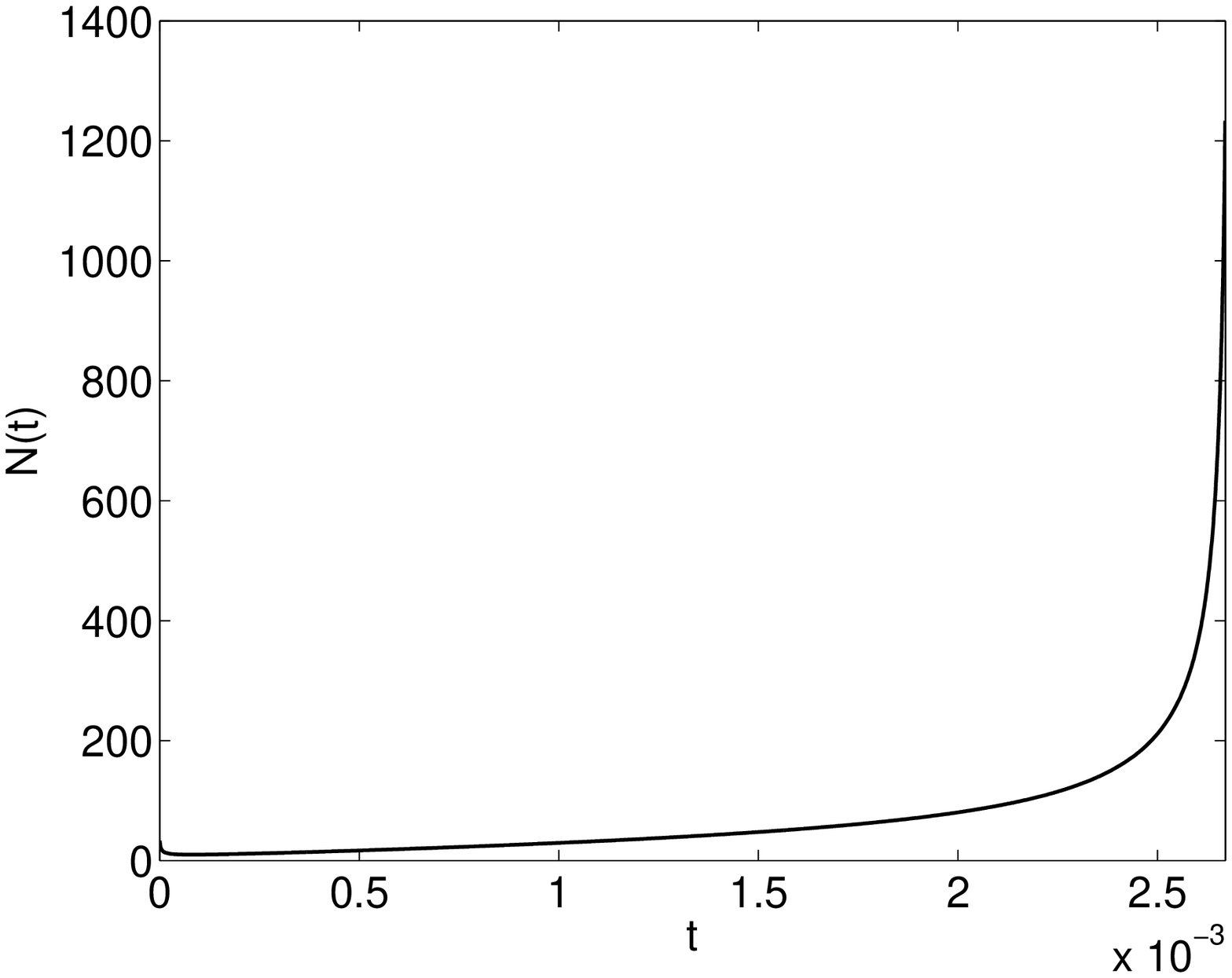}}
\vspace{-5mm}
\caption{Parameter values are  $a=1$ and $b=0.5$ and this corresponds to a single steady state.
Left: Evolution of the distribution function $p(v,t)$ in time of
an initial Maxwellian centered at $v=1.83$ and with variance $0.003$.
Right: Time evolution of the firing rate; again this seems to be a typical blow-up behaviour.}
\label{blowup05}\end{figure}

{\em Blow up.-} According to our blow-up Theorem \ref{th:bu}, the
blow-up in finite time of the solution happens for any value of
$b>0$ if the initial data is concentrated enough on the firing
rate. In Figures \ref{blowup} and \ref{blowup05}, we show the
evolution on time of the firing rate with an initial data with
mass concentrated close to $V_F$ for values of $b$ in which there
are either a unique or two stationary states. The firing rate
increases without bound up to the computing time. It seems that
the blow-up condition in Theorem \ref{th:bu} is not as restrictive
as to say that the initial data is close to a Dirac Delta at
$V_F$. Let us finally mention that blow-up appears numerically
also in case of $a(N)=a_0+a_1 N$, but here the blow-up scenario is
characterized by a break-up of the condition under which
\eqref{eq:nif3} has a unique solution $N$, i.e.,
$$
a_1\left|\frac{\partial p}{\partial v}(V_F,t)\right|<1  \, .
$$
Therefore, the blow-up in the value of the firing rate appears
even if the derivative of $p$ at the firing voltage does not
diverge.

\section{Conclusion}
\label{sec:conc}

The nonlinear noisy leaky  integrate and fire (NNLIF) model is a
standard Fokker-Planck equation describing spiking events in
neuron networks. It was observed numerically in various places,
but never stated as such, that a blow-up phenomena can occur in
finite time. We have described a class of situations where we can
prove that this happens. Remarkably, the system can blow-up for
all connectivity parameter $b>0$, whatever is the (stabilizing)
noise.

The nature of this blow-up is not mathematically proved.
Nevertheless, our estimates in Lemma \ref{lm:apriori} indicate
that it should not come from a vanishing behaviour for $v \approx
-\infty$, or a lack of fast decay rate because the second moment
in $v$ is controlled uniformly in blow-up situations.
Additionally, numerical evidence is that the firing rate $N(t)$
blows-up in finite time whenever a singularity in the system
occurs. This scenario is compatible with all our theoretical
knowledge on the NNLIF and in particular with $L^1$ estimates on
the total network activity (firing rate $N(t)$).

Blow-up has also been proved to occur in the deterministic
quadratic adaptive Integrate-and-Fire model in
\cite{Touboul_AQIF}. The blow-up scenario here is quite different
from ours in many aspects; the model is a linear model (in our
terminology) for a single neuron not a network.  Remarkably, the
blow-up scenario arises on the adaptation variable that we do not
have in the NNLIF.  These are interesting questions to know if
blow-up could occur in a noisy 'linear' adaptive situation.

We have established that the set of steady states can be empty, a
single state or two states depending on the network connectivity.
These are all compatible with blow-up profile, and when they
exist, numerics can exhibit convergence. Several questions are
left open; is it possible to have triple or more steady states?
Which of them are stable? Can a bifurcation analysis help to
understand and compute the set of steady states?

%
%

\bigskip
\footnotesize \noindent\textit{Acknowledgments.} The first two
authors acknowledge support from the project MTM2008-06349-C03-03
DGI-MCI (Spain). JAC acknowledges support from the 2009-SGR-345
from AGAUR-Generalitat de Catalunya. BP has been supported by the
ANR project MANDy, Mathematical Analysis of Neuronal Dynamics,
ANR-09-BLAN-0008-01. The three authors thank the CRM-Barcelona and
Isaac Newton Institute where this work was started and completed
respectively.


\begin{thebibliography}{99}

\bibitem{brunel}N. Brunel, Dynamics of sparsely connected networks of excitatory
and inhibitory spiking networks, J. Comp. Neurosci. 8, (2000)
183--208.

\bibitem{BDP}
A. Blanchet, J. Dolbeault, and B. Perthame, Two-dimensional
{K}eller-{S}egel model: optimal critical mass and qualitative
properties of the solutions, Electron. J. Differential Equations
44, (2006) 32 pp. (electronic).

\bibitem{BrGe} R. Brette and W. Gerstner, Adaptive exponential integrate-and-fire model as an effective description of neural activity. Journal of neurophysiology, 94, (2005) 3637--3642.

\bibitem{BrHa}
N. Brunel and V. Hakim, Fast global oscillations in networks of
integrate-and-fire neurons with long fiting rates, Neural
Computation 11, (1999) 1621--1671.

\bibitem{BCD}
C. Buet, S. Cordier, and V. Dos Santos, A conservative and entropy
scheme for a simplified model of granular media, Transp. Theory
Statist. Phys. 33, (2004), 125--155.

\bibitem{CCTa}
M. J. C\'aceres, J. A. Carrillo, and L. Tao,
A numerical solver for a nonlinear Fokker-Planck equation
representation of neuronal network dynamics, to appear in J. Comp.
Phys.

\bibitem{CBGW}
A. Compte, N. Brunel, P. S. Goldman-Rakic, and X.-J. Wang,
Synaptic mechanisms and network dynamics underlying spatial
working memory in a cortical network model, Cerebral Cortex 10,
(2000) 910--923.

\bibitem{CPZ}
L. Corrias, B. Perthame, and H. Zaag, Global solutions of some
chemotaxis and angiogenesis systems in high space dimensions,
Milan J. Math. 72, (2004) 1--28.

\bibitem{GK} W. Gerstner and W. Kistler. Spiking neuron models. Cambridge Univ. Press (2002).

\bibitem{GG09} M. d. M. Gonz\'alez and M. P. Gualdani, Asymptotics for a symmetric
equation in price formation, App. Math. Optim. 59, (2009)
233--246.

\bibitem{G} T. Guillamon, An introduction to the mathematics of
neural activity, Butl. Soc. Catalana Mat. 19 (2004), 25--45.

\bibitem{lapicque} L. Lapicque, Recherches quantitatives sur l'excitation
\'electrique des nerfs trait\'ee comme une polarisation, J.
Physiol. Pathol. Gen. 9, (1907) 620--635.

\bibitem {ledoux} M. Ledoux, The concentration of measure phenomenon.
AMS math. surveys and monographs 89. Providence 2001.

\bibitem{mg} M. Mattia and P. Del Giudice,
Population dynamics of interacting spiking neurons, Phys. Rev. E
66, (2002) 051917.

\bibitem{RefMMP} P. Michel, S. Mischler, and B. Perthame,
General relative entropy inequality: an illustration on growth
models, J.Math.Pures Appl. 84, (2005) 1235--1260.

\bibitem{NKKRC10}
K. Newhall, G. Kova{\v{c}}i{\v{c}}, P. Kramer, A.V. Rangan, and D.
Cai, Cascade-Induced Synchrony in Stochastically-Driven Neuronal
Networks, preprint 2010.

\bibitem{NKKZRC10}
K. Newhall, G. Kova{\v{c}}i{\v{c}}, P. Kramer, D. Zhou, A.V.
Rangan, and D. Cai, Dynamics of Current-Based, Poisson Driven,
Integrate-and-Fire Neuronal Networks, Comm. in Math. Sci. 8 (2010)
541--600.

\bibitem{PPD} K. Pakdaman, B. Perthame, D. Salort,
Dynamics of a structured neuron population, Nonlinearity 23 (2010) 55--75.

\bibitem{PPCV} J. Pham, K. Pakdaman, J. Champagnat and J.-F. Vivert, Activity in sparsely
connected excitatory neural networks: effect of connectivity. Neural Networks 11 (1998) 415--434.

\bibitem{RBW}
A. Renart, N. Brunel, and X.-J. Wang, Mean-Field Theory of
Irregularly Spiking Neuronal Populations and Working Memory in
Recurrent Cortical Networks, Chapter 15 in Computational
Neuroscience: A comprehensive approach, edited by Jianfeng Feng,
Chapman \& Hall/CRC Mathematical Biology and Medicine Series,
2004.

\bibitem{S} C.-W. Shu, Essentially Non-Oscillatory and Weighted Esentially Non-Oscillatory
schemes for hyperbolic conservation laws, Advanced Numerical
Approximation of Nonlinear Hyperbolic Equations, B. Cockburn, C.
Johnson, C.-W. Shu and E. Tadmor (Editor: A. Quarteroni), Lecture
Notes in Mathematics, volume 1697, Springer, 1998, pp. 325-432.

\bibitem{Touboul_AQIF} J. Touboul, Importance of the cutoff value in the
quadratic adaptive integrate-and-fire model. Neural Computation 21, 2114-2122 (2009).

\bibitem{T} H.C. Tuckwell, Introduction to Theoretical
Neurobiology, Cambridge, Cambridge University Press, 1988.

\end{thebibliography}
\end{document}